\documentclass[a4paper,12pt]{article}
\pdfoutput=1
\usepackage{epsfig,graphicx,xcolor,amsbsy,amssymb,latexsym,amsfonts,amsmath,,tcolorbox,setspace}
\usepackage{pstricks}
\usepackage{color}
\usepackage{soul}
\usepackage{placeins}
\usepackage{wrapfig,framed,caption}
\usepackage[makeroom]{cancel}

\usepackage{wasysym}

\usepackage{multirow}

\definecolor{shadecolor}{gray}{0.925}

\usepackage{eurosym}
\usepackage[a4paper]{geometry}
\usepackage{hyperref}
\usepackage{graphicx}
\usepackage{tikz}
\usetikzlibrary{decorations.pathreplacing,decorations.markings,snakes}
\usetikzlibrary{decorations.pathmorphing}
\usetikzlibrary{patterns}
\usetikzlibrary{calc}
\usepackage{xcolor}
\usepackage{amsmath,amssymb,amsfonts,pstricks,setspace}
\usepackage[enableskew]{youngtab}

\usepackage[style=numeric-comp,sorting=none,maxbibnames=9,backend=bibtex]{biblatex}
\numberwithin{equation}{section}

\usepackage{wrapfig}

\newcommand{\bea}{\begin{eqnarray}\displaystyle}
\newcommand{\eea}{\end{eqnarray}}

\newcommand{\fhn}[1]{f^{(#1)}_H}
\newcommand{\hhn}[1]{h^{(#1)}_H}

\newcommand{\tah}{\tau_{H}}
\newcommand{\zh}{z_{H}}
\newcommand{\zs}{z_{S}}

\newcommand{\zp}{z_{P}}
\newcommand{\zgam}{z_{\gamma}}

\newcommand{\ang}{\beta}

\newcommand{\ppar}[1]{p_{#1}^{||}}
\newcommand{\pperp}[1]{p_{#1}^{\perp}}

\newcommand{\sa}[1]{{\sigma^{(1)}_{#1}}}
\newcommand{\se}[1]{{\sigma^{(2)}_{#1}}}
\newcommand{\sd}[1]{{\sigma^{(2)}_{#1}}}

\newcommand{\rc}[1]{{r_{#1}}}
\newcommand{\tc}[1]{{t_{#1}}}

\newcommand{\imp}{b}

\newenvironment{linesmall}[1]
  {\arraycolsep=3pt\scriptsize
   \array{#1}}
  {\endarray}

\title{
{\bf Probing Effective Black Hole Deformations}\\[30pt]}

\author{\large \textsc{Stefan~Hohenegger\footnote{\tt s.hohenegger@ipnl.in2p3.fr}}}

\begin{document}

\maketitle
\thispagestyle{empty}

\maketitle
\thispagestyle{empty}
${}$\\
\begin{center}
\renewcommand{\thefootnote}{\fnsymbol{footnote}}\vspace{-0.5cm}
${}^{\footnotemark[1]}$ Universit\'e Claude Bernard Lyon 1, CNRS/IN2P3, IP2I Lyon, UMR 5822, Villeurbanne, F-69100, France\\[3.2cm] 
\end{center}

\begin{abstract}
In recent works \cite{DelPiano:2023fiw,DelPiano:2024gvw}, a framework has been developed to describe (quantum) deformed, spherically symmetric and static black holes in four dimensions. The key idea of this so-called \emph{Effective Metric Description} (EMD) is to parametrise deformations of the classical Schwarzschild geometry by two functions that depend on a physical quantity and which are calculated in a self-consistent way as series expansions in the vicinity of the horizon. In this work we further strengthen this framework by first demonstrating that the corresponding series expansion coefficients can be completely and uniquely determined from measurements that are accessible for observers outside of the event horizon: we propose a Gedankenexperiment, consisting of probes following a free-falling trajectory that send signals to a stationary observer and show how an EMD can be constructed from suitable telemetric data. Furthermore, by linking the expansion coefficients of the EMD to the invariant eigenvalues of the energy momentum tensor, we determine a system of physical fields that provides an effective Einstein equation for the deformed black hole geometry. In the case of a simplified geometry and assuming that the metric deformations are small, we can write the leading order of the physical fields in a closed form in the metric functions. We illustrate our results at the example of the Hayward space-time.
\end{abstract}

\newpage

\tableofcontents

\section{Introduction}
\emph{Black holes} are among the most intriguing and most studied space-time geometries within General Relativity (GR). Theories going beyond GR, in particular by incorporating quantum effects, predict deformations of these classical solutions and considerable efforts have been devoted in recent years to their theoretical study. With the advent of gravitational wave astronomy \cite{LIGOScientific:2016aoc} and direct observations of their immediate surroundings \cite{EventHorizonTelescope:2019dse, EventHorizonTelescope:2021dqv, EventHorizonTelescope:2022wkp, EventHorizonTelescope:2022xqj,Vagnozzi:2022moj}, black holes have also started playing a much more important role in experimental astrophysics. The increasing number of measurements and advances in the sensitivity of the detectors, spark the hope to find hints or indications for theories beyond GR and a potential theory of Quantum Gravity.

In view of the large number of theoretical models and geometries that go beyond the classical solutions of GR (see for example~\cite{bardeen1968,Dymnikova:1992ux,Dymnikova:2004qg,Garcia95,Donoghue:2001qc,Bjerrum-Bohr:2002fji,Hayward:2005gi,Kirilin:2006en,Frolov:2016pav,Calmet:2017qqa,Simpson:2018tsi,Calmet:2019eof,Simpson:2019mud,Nicolini:2019irw,Calmet:2021lny,Battista:2023iyu}), it is difficult to extract common features that can be reliably compared to experimental data. For this reason, in recent works \cite{Binetti:2022xdi,DelPiano:2023fiw,DelPiano:2024gvw} an effective framework has been proposed to describe black hole geometries in a model independent way that can capture universal features and compute physical observables in a way that makes them comparable to experimental data. Focusing for simplicity on spherically symmetric and static space-times, the approach assumes that the black hole can (at least effectively) still be described geometrically: the core of the framework is to formulate deformations of the classical Schwarzschild metric \cite{Schwarzschild:1916uq} by two functions that depend only on a physical quantity, such as a suitable proper distance or curvature invariants. This guarantees not only that the deformed metrics respect the same symmetries as their classical counterparts, but also makes the deformation functions physically meaningful and thus accessible by experiments, \emph{i.e.} they can be probed locally.

The framework advocated in \cite{DelPiano:2023fiw,DelPiano:2024gvw,Binetti:2022xdi} (see also \cite{DAlise:2023hls,DelPiano:2024nrl,Hohenegger:2024kbg,DamiaPaciarini:2025akr,DamiaPaciarini:2025qhq}) are effective metric descriptions (EMD) of the (quantum)\footnote{The motivation in \cite{Binetti:2022xdi,DelPiano:2023fiw,DelPiano:2024gvw} is to describe deformations of the Schwarzschild space-time due to \emph{quantum} corrections going beyond classical GR. However, the developed formalism can also incorporate any other corrections stemming from a theory beyond GR, provided the black hole (and the physics in its gravitational field) is still (effectively) described by a metric and compatible with the symmetries of the original space-time.} deformed black hole:  while the choice of the physical quantity that is used to parametrise the deformation functions of the spherically symmetric and static metric does not matter (in the sense that different choices can unambiguously be mapped into each other \cite{DelPiano:2024gvw}), it needs to be computed in a self-consistent fashion. This is indeed non-trivial, since viable physical quantities generally depend on the metric itself, which therefore generally leads to complicated, non-linear consistency conditions. These have been solved in \cite{DelPiano:2023fiw} by using a(n infinite) series expansion of the metric deformations in the vicinity of the black hole horizon. In this way, the physical input that characterises a particular black hole model, is the set of expansion coefficients appearing in these series. Physical observables are captured in a universal fashion, by writing them in terms of (a subset of) these coefficients.

Although these series expansions are expected to only have a finite radius of convergence, it was demonstrated in \cite{DelPiano:2024nrl}, how to extend this framework to larger regions of space-time by using Padé approximants of the metric deformations (or suitable observables). In this way, in particular the black hole shadow was studied as an example of a physical observable, for which an effective approximation was found, linking it directly to the metric deformations. Furthermore, non-trivial conditions for the physical expansion coefficients were found in \cite{DAlise:2023hls}, by imposing (reasonable) positivity conditions asymptotically far away from the black hole. Similarly, in \cite{Li:2023djs} the EMD has been used to describe the precession of bound orbits of the black hole. Finally, we also note that the framework of EMDs has recently been extended to black holes in three dimensions \cite{Hohenegger:2024kbg} that are deformations of the Bañados--Teitelboim--Zanelli (BTZ) black hole \cite{Banados:1992wn, Banados:1992gq, Bambi:2023try}, as well as charged (four-dimensional) black holes \cite{DamiaPaciarini:2025akr} that are deformations of the Reissner-Nordstr\"om geometry \cite{Reissner,Weyl,Nordstrom,Baker}.

A core aspect of the EMD-framework is that the input coefficients (which enter through the series expansions of the deformation functions at the vicinity of the horizon) are physically meaningful, since they encode an expansion in terms of a physical quantity. Therefore, they (and thus by extension the entire metric deformation) should be accessible by physical measurements from outside of the event horizon of the black hole. The main goal of this article is to explicitly demonstrate this in the case of a spherically symmetric and static black hole. Indeed, in the first part of this work, we shall provide a setup for a Gedankenexperiment that can be entirely realised by observers outside of the black hole and which captures all of the deformation functions. It therefore allows to unambiguously reconstruct the metric deformations. Concretely, we consider two measurements by a spacecraft, capable of holding a fixed (stationary) distance or a circular orbit around the black hole, which releases probes that follow a free-falling radial trajectory. We assume that these probe can send light signals and information back to the spacecraft:
\begin{itemize}
\item[\emph{(i)}] {\bf redshift factor:} in a first measurement, a probe sends a photon along a radial outward trajectory towards the spacecraft, together with information about its frequency and the proper time at which it was emitted. By measuring the frequency of the photon, an observer on the spacecraft can reconstruct a signal $S_1$, corresponding to the redshift factor as a function of the proper time (of the probe). See also \cite{Herrera-Aguilar:2015kea,Becerril:2016qxf,Herrera-Aguilar:2018uxd,Banerjee:2022him} for a similar proposal to measure (the mass) parameters of a Kerr-black hole (and subsequent applications to other geometries \cite{ShankarKuniyal:2017sxc,Uniyal:2017yll,Sheoran:2017dwb,Kraniotis:2019ked,Becerril:2020fek,Morales-Herrera:2024zbr}).
\item[\emph{(ii)}] {\bf critical emission angle:} in a second measurement, a probe sends out photons at an angle (relative to the radial direction) together with information about the proper time when and the angle under which the photon has been emitted. Photons emitted under too large an angle cannot reach the spacecraft, but fall back into the black hole or, in the critical case, orbit around it. We consider as a second signal $S_2$, the largest angle of emitted photons capable of reaching the spacecraft, at a given proper time at the probe.
\end{itemize}
We assume that the telemetry data can be received up to the point where the probes cross the event horizon of the black hole.\footnote{Since signals emitted from the event horion take infinitely long to reach the spacecraft, this technically means that we assume an infinitely long observation time. However, in practice we only assume a sufficiently long observation time, such that the asymptotics of the telemetry data can be safely extrapolated to the point when the probe crosses the horizon.} We demonstrate, that the signals $S_{1,2}$ received in this way allow to fully reconstruct the black hole geometry outside of the event horizon. Concretely, using an EMD based on the proper time of a probe to hit the horizon, we show how to iteratively express all the expansion coefficients of the metric deformation functions in terms of (expansion coefficients) of the telemetry. This shows that the EMD can indeed uniquely be determined by physical measurements and also underlines the importance of an EMD as a physical description of the deformed metric. In order to make contact to previous results in the literature, we provide an explicit relation to an EMD based on the proper distance to the horizon and express the Hawking temperature in terms of (expansion coefficients of) the telemetry signals. To illustrate our computations, we showcase our results for the Hayward black hole geometry~\cite{Hayward_2006}, which is a non-trivial example of a deformed, static, spherically symmetric black hole.

In order to get a better understanding of an effective physical system that can reproduce the deformed black hole geometries probed in the first part of this work, we make contact to the energy momentum tensor $T_{\mu\nu}$. The latter can be calculated from the black hole geometry: indeed, we show that the invariant eigenvalues of the energy momentum can also serve as telemetry data to reconstruct the physical metric deformations. This means, a falling probe capable of measuring these eigenvalues would also be capable of iteratively reconstructing the physical expansion coefficients in the metric. More importantly, although calculated from purely geometric quantities, $T_{\mu\nu}$ can also be replicated by a physical system made of local fields. Although in general not unique, a minimal such realisation (in the static and spherically symmetric case) was provided in \cite{Boonserm:2015aqa}, in the form of a massless scalar field, an electric-like field and a perfect fluid. We express these fields in terms of the expansion coefficients of the metric deformations, thus making a first step towards an effective field equation for the deformed metrics. Furthermore, assuming the corrections of the metric to be small (in the same manner as in \cite{Rezzolla:2014mua,DelPiano:2024nrl}), which is expected if these corrections are due to (quantum) effects beyond GR, we provide for a simplified geometry, closed expressions for the invariant eigenvalues of the energy momentum tensor (and thus the physical fields that realise it), corrected to leading order. These are formulated directly in terms of the metric deformations. We showcase the accuracy of these effective field descriptions at the example of the Hayward black hole and find good agreement with numerical computations.

This paper is organised as follows: In Section~\ref{Sect:SignalsTelemetry} we introduce the telemetry signals $S_{1,2}$ and explain how they can be used to reconstruct the physical metric deformation functions. We also express our results in terms of an EMD based on the proper distance to the black hole horizon (which was used in previous works) and illustrate our results at examples. In Section~\ref{Sect:EnergyMomentum}, we relate the invariant eigenvalues of the energy momentum tensor to the physical coefficients of the metric deformation functions. Using the minimal field realisation in \cite{Boonserm:2015aqa}, this therefore provides an effective physical description of the metric deformations. Assuming these deformations to be small in a well-defined manner, allows us to write closed form expressions for the leading order corrections. Finally, Section~\ref{Sect:Conclusions} contains our conclusions. This paper is supplemented by three appendices, which contain additional discussions and computations for the geodesic motion of particles around the black hole and details on certain series expansions.

\section{Metric Deformations from Probe Telemetry}\label{Sect:SignalsTelemetry}
In this Section, we propose a Gedankenexperiment  to probe metric deformations of static, spherically symmetric black hole geometries with a simple horizon. Concretely, we shall explain how to access the physical deformation parameters that enter into the \emph{effective metric descriptions} (EMD) introduced in \cite{DelPiano:2024gvw}. We stress that our discussion is agnostic of the mechanism that is behind the deformation, \emph{i.e.} if the geometry is a solution to some model of quantum gravity or some other model beyond General Relativity (GR): we only assume that the black hole (and the physics in its gravitational field) is still (at least effectively) described geometrically, \emph{i.e.} by a (deformed) metric.  
\subsection{Metric Deformations}
We start by describing the metric deformations. The geometry of a generic, four dimensional spherically-symmetric and static black hole in four dimensions (with Lorentzian signature) is given by the following metric \cite{Carroll:2004st,Misner:1973prb,penrose2005reality,hartle2003gravity}
\begin{align}
&ds^2= g_{\mu\nu}\,dx^\mu\,dx^\nu=-h(z)\,dt^2+\frac{dz^2}{f(z)}+z^2\,d\theta^2+z^2\,\sin^2\theta\,d\varphi^2\,,
\label{SchwarzschildMetric}
\end{align}
with $x^\mu=(x^t,x^z,x^\theta,x^\varphi)^\mu=(t,z,\theta,\varphi)^\mu$. Following the notation and conventions of \cite{DelPiano:2023fiw}, we use units such that $c=1$ and the line element $ds$ is written in a dimensionless form. Furthermore, we write the metric functions $f$ and $h$ as
\begin{align}
&f(z)=1-\frac{\Phi(\tau)}{z}\,,&&\text{and} &&h(z)=1-\frac{\Psi(\tau)}{z}\,.\label{DefDeformation}
\end{align}
Here $\Phi(\tau)$ and $\Psi(\tau)$ parametrise the deformation from the Schwarzschild geometry\footnote{The Schwarzschild black hole \cite{Schwarzschild:1916uq} (which is a solution of GR for a point like massive source), is recovered by the choice $\Phi(\tau)=\Psi(\tau)=2\chi=\text{const.}$, where $\chi$ is the mass of the black hole (in dimensionless units).} and define an EMD. Following the framework outlined in \cite{DelPiano:2024gvw,DelPiano:2023fiw,Binetti:2022xdi}, we shall assume $\Phi$ and $\Psi$ to be functions of a physical quantity, which guarantees that (\ref{SchwarzschildMetric}) has the same symmetries as the classical Schwarzschild geometry. The choice of this physical quantity can be adapted to the concrete physical problem under consideration and in the following we shall choose it to be the proper time $\tau$ of an observer that is freely falling towards the center of the black hole, following a radial, time-like geodesic (see (\ref{DiffEqProperTime}) in Appendix~\ref{App:GeodesicProperTime}). It has recently been argued in \cite{DamiaPaciarini:2025akr} that $\tau$ is a suitable observable to describe deformations of charged four-dimensional black holes.

Furthermore, we shall assume that the functions $f$ and $h$ in (\ref{DefDeformation}) have (simple)\footnote{Our approach can be generalised in a straight-forward manner to allow for higher order zeroes, which is for example relevant to describe deformations of extremal Reissner-Nordstr\"om black holes.} zeroes at the position of the event horizon $\zh$ of the black hole, \emph{i.e.}
\begin{align}
&f(z=\zh)=0=h(z=\zh)\,,&&\text{but} &&\begin{array}{l}f'(z=\zh)=\fhn{1}\neq 0\,,\\[2pt] h'(z=\zh)=\hhn{1}\neq 0\,.\end{array}
\end{align}
We shall also assume $f(z)>0$ and $h(z)>0$ for $z>\zh$, \emph{i.e.} $\zh$ denotes the position of the outermost horizon of the geometry.

\subsection{Series Expansion and Free-Falling Observer}\label{Sect:FreeFallingObserver}
The defining equation of the proper time $\tau$ in (\ref{DiffEqTime}) depends on $f$ and $h$, which through (\ref{DefDeformation}) depend on $\tau$ itself. Thus (\ref{DiffEqTime}) is in fact a non-linear differential equation. Moreover, in the following we propose to determine the deformation functions $\Phi$ and $\Psi$ by observing simple physical processes, such that the former are in fact undetermined in (\ref{DefDeformation}). Instead, we shall determine these functions in a self-consistent way from observations outside of the event horizon.
\begin{figure}[htbp]
\begin{center}
\begin{tikzpicture}
\draw[dashed] (0,0) -- (8,0);
\node at (0,0) {$\bullet$};
\node at (0,0.3) {\footnotesize center of BH};
\node at (8,0) {$\bullet$};
\node at (8,0.3) {\footnotesize spacecraft};
\draw [domain=-20:20] plot ({2*cos(\x)}, {2*sin(\x)});
\node at (2.6,0.75) {\footnotesize horizon};
\node at (4,0.3) {\footnotesize probe};
\draw[thick,red] (4,0) -- (4.1,0);
\draw[->,thick] (4,0) -- (3.5,0); 
\node at (4,0) {$\bullet$};
\node at (0,-0.3) {\footnotesize $0$};
\node at (2.25,-0.2) {\footnotesize $\zh$};
\node at (4,-0.3) {\footnotesize $\zp$};
\node at (8,-0.3) {\footnotesize $\zs$};
\end{tikzpicture}
\end{center}\caption{\emph{Schematic depiction of a spacecraft (at $\zs$) releases a probe (with position $\zp$) that is freely falling towards the center of the black hole, whose horizon is located at $\zh$.
}}
\label{Fig:ProbeFalling}
\end{figure}
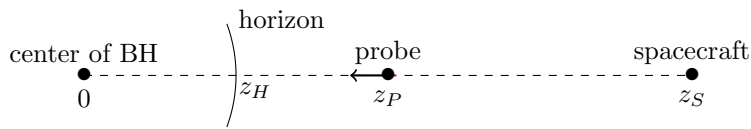
Concretely, as schematically shown in Figure~\ref{Fig:ProbeFalling}, we consider a spacecraft at a fixed position $\zs$ outside of the horizon of the black hole, with the following 4-velocity
\begin{align}
&u_S^\mu=\left(\frac{1}{\sqrt{h(\zs)}},0,0,0\right)\,.\label{FourVelSpacecraft}
\end{align}
We shall not assume to know $\zs$ (or \emph{e.g.} the proper distance of the spacecraft to the horizon of the black hole). We only assume that the spacecraft is capable of holding a fixed position (sufficiently far) outside of the horizon of the black hole. The factor $\sqrt{h(\zs)}$ can be determined from the redshift factor of a photon emitted (with known frequency) to spatial infinity (\emph{i.e.} to earth) and we shall also consider it to be known in the following.

The spacecraft releases a probe that is falling towards the center of the black hole, along a radial time-like geodesic. Upon release from the spacecraft, the probe starts an internal chronometer, whose proper time $\tau$ we shall use to parametrise its position $\zp(\tau)$ (in terms of Schwarzschild coordinates). Concretely, the geodesic of the probe can be described as in Appendix~\ref{App:TimeLikeGeodesic} and the 4-velocity of the probe becomes
\begin{align}
u_P^\mu=\left(\frac{1}{h(\zp)}\,,-\sqrt{\frac{f(\zp)}{h(\zp)}(1-h(\zp))}\,,0\,,0\right)\,,\label{ProbeVelGeo}
\end{align}
where the sign in $u_P^z$ was chosen to correspond to a massive particle falling into the black hole. We also assume that the probe is capable of continuously communicating the reading of its chronometer (and other telemetric data) to the spacecraft by sending light signals.\footnote{We shall study in detail the emission of photons by the probe in the following Subsections.} 

The equation (\ref{DiffEqProperTime}) for the proper time can be re-cast into the following form
\begin{align}
(\zp-\Phi)\,\Psi=\left(\frac{d\zp}{d\tau}\right)^2\,\zp\,(\zp-\Psi)\,,\label{DiffEqExpand}
\end{align}
which provides a relation between the deformation functions $\Phi$ and $\Psi$ and the (unknown) position $\zp$ of the probe. Furthermore, we shall denote the (proper) time at which the probe crosses the horizon as $\tah$ (see (\ref{DiffEqTime})). The spacecraft can determine $\tah$ as the last (asymptotically) transmitted proper time by the probe, since after having passed the horizon no further signals can reach the spacecraft. Although, as we shall see, the last transmission of the probe reaches the spacecraft after an infinitely long period, in the following we shall assume that the spacecraft can observe the signals of the probe for a sufficiently long time, that the asymptotic telemetry can at least be reliably extrapolated.

Following the framework advocated in \cite{DelPiano:2023fiw,Binetti:2022xdi}, we search for a solution of (\ref{DefDeformation}) by expanding close to the horizon. To this end, we define the following series 
\begin{align}
&\zp=\zh+\sum_{n=1}^\infty \alpha_n\, x^n\,,&&\Phi=\zh+\sum_{n=1}^\infty \varphi_n\, x^n\,,&&\Psi=\zh+\sum_{n=1}^\infty \psi_n\, x^n\,,&&\text{with} &&x:=\tah-\tau\,,\label{SeriesExpansionsDeformations}
\end{align}
which we assume to be convergent at least for sufficiently small values of $x$. We consider the expansion coefficients $\{\alpha_n\}$, $\{\varphi_n\}$ and $\{\psi_n\}$ as well as $\zh$ to be unknown. Inserting these series expansions into (\ref{DiffEqExpand}), however, allows to find infinitely many relations among these coefficients:
{\allowdisplaybreaks
\begin{align}
\sum_{p=1}^\infty \zh &(\alpha_p-\varphi_p)x^p+\sum_{p=2}^\infty x^p\sum_{n=1}^{p-1}(\alpha_n-\varphi_n)\psi_{p-n}\nonumber\\
&=\sum_{p=1}^\infty x^p\sum_{r=0}^{p-1}\sum_{n=1}^{r+1}n(r-n+2)\zh\alpha_n\alpha_{r-n+2}(\alpha_{p-r}-\psi_{p-r})\nonumber\\
&\hspace{0.5cm}+\sum_{p=2}^\infty x^p\sum_{r=0}^{p-2}\sum_{n=1}^{r+1}\sum_{\ell=1}^{p-r-1} n(r-n+2)\alpha_n\alpha_{r-n+2}\alpha_{p-r-\ell}(\alpha_\ell-\psi_\ell)\,.\label{MasterSeriesProperTime}
\end{align}
}
Matching the terms of order $\mathcal{O}(x^1)$ on both sides, we require
\begin{align}
(\alpha_1-\varphi_1)=\alpha_1^2(\alpha_{1}-\psi_{1})\,,\label{RelAlpha1}
\end{align}
which fixes $\alpha_1$ as a function of $\varphi_1$ and $\psi_1$.\footnote{Notice that $\varphi_1=\psi_1$ leads to the condition $\alpha_1=\pm 1$ and $\psi_1\neq \alpha_1\neq\varphi_1$ in general.} In general, (\ref{RelAlpha1}) is a cubic equation in $\alpha_1$, however, not all solutions lead to a smooth black hole geometry. Indeed, as discussed in Appendix~\ref{Sect:CurvatureInvariants}, demanding finiteness of curvature invariants (notably the Ricci- and Kretschmann scalar) at the horizon (\emph{i.e.} and thus the absence of a curvature singularity), leads to the stronger equation (\ref{SectStrongerEq}), which is compatible with (\ref{RelAlpha1}). From the terms of order $\mathcal{O}(x^p)$ in (\ref{MasterSeriesProperTime}) with $p\geq 2$ we obtain
\begin{align}
\zh(\alpha_p-\varphi_p)+&\sum_{n=1}^{p-1}(\alpha_n-\varphi_n)\psi_{p-n}=\sum_{r=0}^{p-1}\sum_{n=1}^{r+1}n(r-n+2)\zh\alpha_n\alpha_{r-n+2}(\alpha_{p-r}-\psi_{p-r})\nonumber\\
&+\sum_{r=0}^{p-2}\sum_{n=1}^{r+1}\sum_{\ell=1}^{p-r-1} n(r-n+2)\alpha_n\alpha_{r-n+2}\alpha_{p-r-\ell}(\alpha_\ell-\psi_\ell)\,.
\end{align}
Arranging all terms that contain $\alpha_p$ on one side, we find 
\begin{align}
\alpha_p&\,\zh\left(1-2p \alpha_{1}(\alpha_{1}-\psi_1)-\alpha_1^2\right)\nonumber\\
&=\zh (\varphi_p-\alpha_1^2\psi_p)-\sum_{n=1}^{p-1}(\alpha_n-\varphi_n)\psi_{p-n}+\sum_{n=2}^{p-1}n(p-n+1)\zh\alpha_n\alpha_{p-n+1}(\alpha_{1}-\psi_{1})\nonumber\\
&\hspace{0.5cm}+\sum_{r=1}^{p-2}\sum_{n=1}^{r+1}n(r-n+2)\zh\alpha_n\alpha_{r-n+2}(\alpha_{p-r}-\psi_{p-r})\nonumber\\
&\hspace{0.5cm}+\sum_{r=0}^{p-2}\sum_{n=1}^{r+1}\sum_{\ell=1}^{p-r-1} n(r-n+2)\alpha_n\alpha_{r-n+2}\alpha_{p-r-\ell}(\alpha_\ell-\psi_\ell)\hspace{2cm}\forall p\geq 2\,.\label{HigherAlphaOrders}
\end{align}
Since all terms on the right hand side of this expression contain only $\alpha_{p'}$ with $p'<p$, this relation allows to iteratively determine all $\alpha_p$ as a function of the $\varphi_n$ and $\psi_n$. Notice in particular that $\alpha_p$ depends on finitely many $\varphi_{p'}$ and $\psi_{p'}$ and is linear in $\varphi_p$ and $\psi_p$ for $p\geq 2$.

\subsection{Transmission of Telemetry}
So far, we have assumed that a probe simply transmits the reading of its internal chronometer to the spacecraft. In order to determine the geometry of the black hole around the horizon, we require additional information. In the following we shall therefore assume that probes (at positions $\zp$ outside of the horizon) may emit photons along with information about their initial conditions that are recovered by the spacecraft.
\subsubsection{Radial Emission and Redshift Factor}
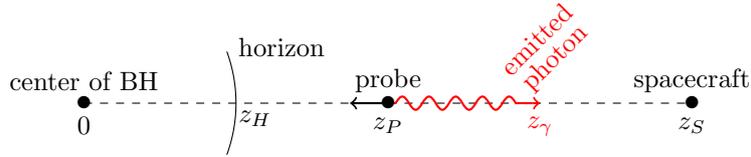
\begin{figure}[htbp]
\begin{center}
\begin{tikzpicture}
\draw[dashed] (0,0) -- (8,0);
\node at (0,0) {$\bullet$};
\node at (0,0.3) {\footnotesize center of BH};
\node at (8,0) {$\bullet$};
\node at (8,0.3) {\footnotesize spacecraft};
\draw [domain=-20:20] plot ({2*cos(\x)}, {2*sin(\x)});
\node at (2.6,0.75) {\footnotesize horizon};
\node at (4,0.3) {\footnotesize probe};
\draw[thick,red] (4,0) -- (4.1,0);
\draw[snake=coil, segment aspect=0,thick,red] (4.1,0) -- (5.9,0);
\draw[->,thick,red] (5.9,0) -- (6,0);
\node[red,rotate=45] at (5.95,0.85) {\footnotesize emitted};
\node[red,rotate=45] at (6.2,0.6) {\footnotesize photon};
\draw[->,thick] (4,0) -- (3.5,0); 
\node at (4,0) {$\bullet$};
\node at (0,-0.3) {\footnotesize $0$};
\node at (2.25,-0.2) {\footnotesize $\zh$};
\node at (4,-0.3) {\footnotesize $\zp$};
\node[red] at (6,-0.3) {\footnotesize $\zgam$};
\node at (8,-0.3) {\footnotesize $\zs$};
\end{tikzpicture}
\end{center}
\caption{\emph{Schematic depiction of the free-falling probe $\zp$ sending a photon back to the spacecraft (at $\zs$). The photon is following a light-like radial trajectory with radial position $\zgam$. 
}}
\label{Fig:RadialPhoton}
\end{figure}

We first assume that a photon is emitted from a free-falling probe radially outwards towards the spacecraft, as schematically shown in Figure~\ref{Fig:RadialPhoton}.
The radial position of the photon shall be denoted $\zgam$ and the corresponding 4-velocity $u_\gamma^\mu$ can be obtained as described in Appendix~\ref{App:LightLikeGeodesic} for $L=0$:
\begin{align}
u_\gamma^\mu=\left(\frac{E_\gamma}{h(\zgam)}\,,\sqrt{\frac{f(\zgam)}{h(\zgam)}}\,E_\gamma\,,0\,,0\right)\,.
\end{align}
We recall that the frequency of a photon emitted or detected by an observer with 4-velocity $v^\mu$ is given by (see \emph{e.g.} \cite{Herrera-Aguilar:2012jhe})
\begin{align}
\omega=-g_{\mu\nu}\,v^\mu\,u_\gamma^\nu\,.\label{DefFrequencyDetect}
\end{align}
We can therefore determine the frequencies of the photon observed upon emission by the probe and upon detection by the spacecraft
\begin{enumerate}
\item[\emph{(i)}] {\bf emission:} the photon is emitted at $\zgam=\zp$ and an observer co-moving with the probe registers the emission frequency (\ref{DefFrequencyDetect}) for $v^\mu=u_P^\mu$ given in (\ref{ProbeVelGeo}):
\begin{align}
\omega_e=h(\zp)\,u^t_P\,u^t_\gamma-\frac{1}{f(\zp)}\,u_P^z\,u_\gamma^z=\frac{E_\gamma}{h(\zp)}\,\left(1+\sqrt{1-h(\zp)}\right)\,.\label{EmissionFrquency}
\end{align}
We assume that this frequency is continuously transmitted to the spacecraft as a function of the proper time of the probe. 
\item[\emph{(ii)}] {\bf detection:} the photon is detected by the spacecraft at $\zgam=\zs$, which measures the detection frequency (\ref{DefFrequencyDetect}) for $v^\mu=u_S^\mu$ given in (\ref{FourVelSpacecraft}):
\begin{align}
\omega_d=h(\zs)\,u_S^t\,u_\gamma^t=\frac{E_\gamma}{\sqrt{h(\zs)}}\,.
\end{align} 
\end{enumerate}
The spacecraft therefore registers the following quotient of frequencies 
\begin{align}
S_1=\sqrt{h(\zs)}\,\,\frac{\omega_d}{\omega_e}=\frac{h(\zp)}{1+\sqrt{1-h(\zp)}}\,.\label{Telemetry1}
\end{align}
where, as mentioned before, the factor $\sqrt{h(\zs)}$ can be determined by sending a signal from the spacecraft to an asymptotically distant observer and the value of $\omega_e$ has been transmitted by the probe separately to the spacecraft as a function of the proper time $\tau$ at which the photon has been emitted at $\zp$.

We remark in passing that the (proper) time at which the spacecraft receives the photon is
\begin{align}
\tau^S_d=-\sqrt{h(\zs)}\int_{\zs}^{\zp}\frac{dz}{\sqrt{f(z)\,h(z)}}\left[1+\frac{1}{\sqrt{1-h(z)}}\right]\,.\label{ReceiveTime}
\end{align}
Since both $f$ and $h$ are assumed to have a simple zero for $z=\zh$, (\ref{ReceiveTime}) has a non-integrable singularity for $\zp=\zh$, such that the signal emitted by the probe at the horizon takes an infinitely long time (measured by the spacecraft) to reach the spacecraft. As mentioned before, we shall assume that the spacecraft can observe the fall of the probe for a sufficiently long time (such that signals of the latter emitted sufficiently close to the horizon can be detected), in order to at least extrapolate all relevant telemetry data, such as $S_1$ for small values of $x=\tah-\tau$.
\subsubsection{Equatorial Emission and Critical Angle}\label{Sect:AngleEmission}
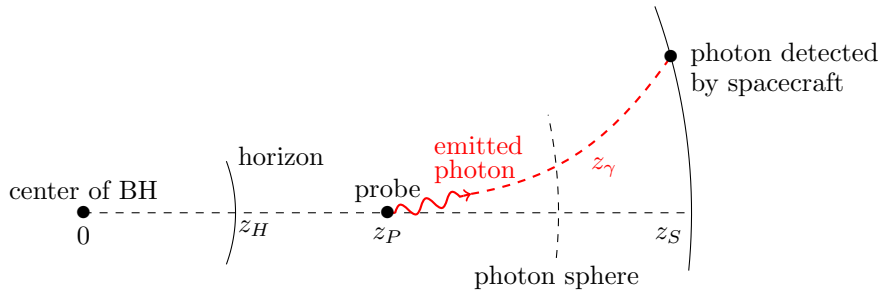
\begin{figure}[htbp]
\begin{center}
\begin{tikzpicture}
\draw[dashed] (0,0) -- (8,0);
\node at (0,0) {$\bullet$};
\node at (0,0.3) {\footnotesize center of BH};
\draw [domain=-20:20] plot ({2*cos(\x)}, {2*sin(\x)});
\node at (2.6,0.75) {\footnotesize horizon};
\node at (4,0.3) {\footnotesize probe};
\draw[thick,red] (4,0) -- (4.1,0);
\draw[snake=coil, segment aspect=0,thick,red,dashed] (5.1,0.25) to [out=10,in=235] (7.72741, 2.07055);
\draw[snake=coil, segment aspect=0,thick,red] (4.1,0) -- (5,0.225);
\draw[thick,red,->] (5,0.225) -- (5.1,0.25);
\node[red] at (5.2,0.9) {\footnotesize emitted};
\node[red] at (5.15,0.55) {\footnotesize photon}; 
\node at (4,0) {$\bullet$};
\node at (0,-0.3) {\footnotesize $0$};
\node at (2.25,-0.2) {\footnotesize $\zh$};
\node at (4,-0.3) {\footnotesize $\zp$};
\node[red] at (6.85,0.6) {\footnotesize $\zgam$};
\node at (7.7,-0.3) {\footnotesize $\zs$};
\draw [domain=-5.5:20] plot ({8*cos(\x)}, {8*sin(\x)});
\draw [domain=-5.5:12,dashed] plot ({6.25*cos(\x)}, {6.25*sin(\x)});
\node at ({8*cos(15)}, {8*sin(15)}) {$\bullet$};
\node at (9.225,2.1) {\footnotesize photon detected};
\node at (9,1.7) {\footnotesize by spacecraft};
\node at (6.225,-0.85) {\footnotesize photon sphere};
\end{tikzpicture}
\end{center}
\caption{\emph{Schematic depiction of the probe located at $\zp$ sending a photon back to the spacecraft (which is at a radial distance $\zs$). The photon, at a radial position $\zgam$, is emitted in the equatorial plane of the black hole, but not along a radial trajectory, as shown by the dashed red line. 
}}
\label{Fig:AnglePhoton}
\end{figure}

We next consider that a probe, located at $\zp$, emits a photon in the equatorial plane of the black hole (\emph{i.e.} for $\theta=\pi/2$), but at a non-trivial angle relative to the radial direction, as shown in Figure~\ref{Fig:AnglePhoton}. This photon is recovered by the spacecraft at a radial distance $\zs$.\footnote{As before, we shall not require to know $\zs$ explicitly, but only that the spacecraft is capable of keeping its position at a fixed orbit.} 
For the emitted photon, we now consider a 4-velocity  $u^\mu_\gamma=\left(u_\gamma^t,u_\gamma^z,0,u_\gamma^\varphi\right)$ as in (\ref{DefFourVelocityPhoton}) of Appendix~\ref{App:LightLikeGeodesic} with conserved energy $E_\gamma$ and angular momentum $L\neq 0$.
In order to introduce the notion of an angle under which the photon is emitted, we pass to the tangent frame at the position of the probe by introducing the vierbein $e_\mu^a$ (for $a\in\{0,1,2,3\}$), with
\begin{align}
&e_t^0=\sqrt{h(\zp)}\,,&&e_z^1=\frac{1}{\sqrt{f(\zp)}}\,,&&e_\theta^2=\zp\,,&&e_\varphi^3=\zp\,,
\end{align}
and all remaining components vanishing. The photon 4-velocity in the tangent frame therefore becomes
\begin{align}
u_\gamma^a=\left(\frac{E_\gamma}{\sqrt{h(\zp)}}\,,E_\gamma\,\sqrt{\frac{1}{h(\zp)}\left(1-\frac{L^2 h(\zp)}{\zp^2 E_\gamma^2}\right)},0,\frac{L}{\zp}\right)\,,
\end{align}
which satisfies $\eta_{ab}\,u_\gamma^a\,u_\gamma^b=0$. For concreteness, we shall then define the angle from the 4-velocity of the photon relative to the radial direction in the tangent frame\footnote{We remark that this definition of the angle $\beta$ is equivalent to the prescription given in \cite{Synge} for a static observer at $\zp$, which considers the triangle with corners at $\{(\zp,\theta=\tfrac{\pi}{2},\varphi),(\zp+dz,\theta=\tfrac{\pi}{2},\varphi),(\zp,\theta=\tfrac{\pi}{2},\varphi+d\varphi)\}$ in the three-dimensional space with the metric element (\ref{SchwarzschildMetric}) for $dt=0$. The angle $\beta$ is then given through $\cot\beta=\frac{1}{\zp\sqrt{f(\zp)}}\,\frac{u^z_\gamma}{u^\varphi_\gamma}\big|_{z-\zp}$ with $u_\gamma^\mu$ as in (\ref{DefFourVelocityPhoton}), which leads to the same expression for $\sin\beta$ as in (\ref{CosAngle}).}
\begin{align}
&\parbox{3.5cm}{\begin{tikzpicture}
\node at (-0.25,1.8) {{\footnotesize tangent frame:}};
\draw[dashed] (0,0) -- (2,0);
\node at (0,0) {$\bullet$}; 
\node at (-0.25,-0.25) {$\zp$};
\draw[thick,dashed] (0,0) -- (1.5,1.5);
\draw[snake=coil, segment aspect=0,thick,red,->] (0,0) -- (1.5,1.5);
\draw[thick] (1.5,0) -- (1.5,1.5);
\draw[thick] (1.5,0) -- (0,0);
\draw [domain=0:45] plot ({0.9*cos(\x)}, {0.9*sin(\x)});
\node at (0.6,0.25) {$\ang$};
\node at (0.5,1.1) {$u_\gamma^0$};
\node at (1.85,0.75) {$u_\gamma^3$};
\node at (0.9,-0.3) {$u_\gamma^1$};
\end{tikzpicture}
}&&\text{with} &&\sin\ang=\frac{u_\gamma^3}{u_\gamma^0}=\frac{L}{\zp}\,\frac{\sqrt{h(\zp)}}{E_\gamma}\,\,.\label{CosAngle}
\end{align}
We shall assume that the probe is capable of transmitting\footnote{For simplicity, in the definition of the angle, we do not take into account the motion of the probe, \emph{i.e.} we implicitly assume that once the probe reaches $\zp$ it holds its position during the emission of this photon.} the angle $\beta$ to the spacecraft. Together with the frequency of the photon, this allows to express the conserved quantities $E_\gamma$ and $L$ of the photon geodesic. For the further computations, however, we shall only require the combination
\begin{align}
\imp:=\frac{L}{E_\gamma}=\frac{\zp\,\sin\ang}{\sqrt{h(\zp)}}\,,
\end{align}
which depends also on the position $\zp$ of the probe where it emits the photon.  The angle (as seen from the black hole), under which the spacecraft detects the photon at $z=\zs$ can be computed from
\begin{align}
&\frac{d\varphi}{dz}=\frac{u_\gamma^\varphi}{u_\gamma^z}\,,&&\text{such that} &&\varphi=\int_{\zp}^{\zs}\frac{\imp}{z^2}\,\frac{dz}{\sqrt{\frac{f}{h}\,\left(1-\frac{\imp^2 h}{z^2}\right)}}\,.\label{FormAngleSpacecraft}
\end{align}
While the angle $\varphi$ can be measured by the spacecraft, the integral presentation notably requires to know the metric functions $f$ and $h$ along the entire path of the integral. However, (\ref{FormAngleSpacecraft}) offers another way of extracting information about the black hole geometry at $\zp$: indeed the angle $\varphi$ is well-defined only as long as
\begin{align}
&\sin^2\ang< \frac{z^2}{h(z)}\,\frac{h(\zp)}{\zp^2}\,,&&\forall z\in[\zp,\zs]\,,\label{RealCond}
\end{align}
where $U(z):=\frac{h(z)}{z^2}$ is the potential, whose maxima determine circular photon orbits. Here we assume the scenario $\zp<z_{\text{ph}}<\zs$, as depicted in Figure~\ref{Fig:AnglePhoton}, where $z_{\text{ph}}$ denotes the position of the (last) photon sphere (\emph{i.e.} the position of the (last) circular photon orbit around the black hole) outside of the horizon. That is, we assume the spacecraft to be sufficiently far away from the black hole and the probe to have already advanced fairly close to the horizon. For $z=\zp=z_{\text{ph}}$ the condition (\ref{RealCond}) becomes $\sin^2\beta< 1$, where for an emission with $\beta=\pi/2$ (such that $\sin\beta=1$), the photon is captured on a circular orbit around the black hole (and cannot reach the spacecraft in finite time). For $\zp<z_{\text{ph}}$ the condition (\ref{RealCond}) is more constraining and requires the emission of the photon below a critical angle $\ang_c(\zp)<\tfrac{\pi}{2}$. Indeed, for $\tfrac{\pi}{2}\geq\ang\geq \ang_c(\zp)$, a photon emitted at $\zp$ cannot reach the spacecraft, but instead gets captured on a closed orbit around the black hole (at $z_{\text{ph}}$). We assume that the spacecraft can measure the critical angle $\ang_c$ as a function of $x$. We then define the telemetry signal 
\begin{align}
S_2:=\frac{h(z_{\text{ph}})}{z_{\text{ph}}^2}\,\sin^2\ang_c=\frac{\sin^2\ang_c}{b_{\text{sh}}^2}
=\frac{h(\zp)}{\zp^2}\,.\label{Telemetry3}
\end{align}
Here $b_{\text{sh}}^2=U^{-1}(z_{\text{ph}})$ is the black hole shadow, which can be measured by an asymptotically distant observer (see \cite{Synge,Perlick:2021aok,Cunha:2020azh} and also \cite{DelPiano:2024nrl}) through the minimal impact parameter of photon trajectories bending around the black hole. We shall therefore assume $b_{\text{sh}}$ to be known, such that $S_2$ can be fully determined by measurements from outside of the black hole.

We remark that in principle multiple (stable and unstable) photon spheres may exist outside of the black hole (see \emph{e.g.} \cite{Gan:2021pwu,Cunha:2020azh}): in writing (\ref{Telemetry3}) we have used that the relevant constraints for the detection of the photon by the spacecraft are imposed by the same photon sphere that is responsible for the creation of the black hole shadow (\emph{i.e.} generally the outermost). However, not least because the critical angle as the maximal emission angle under which the photon can still reach the spacecraft, might be difficult to determine concretely (see \emph{e.g.} \cite{Gralla:2019xty} for subtleties in defining the shadow), we have also explored other telemetries than (\ref{Telemetry3}): for completeness we explain in Appendix~\ref{App:FullAngleExpansion}, how the detection of photons at any angle (and its variation) can still be used to extract information of the geometry at the point of emission by the probe.

\subsection{Series Expansions and Self-Consistent Solution}
We shall now demonstrate that the reception of the telemetry data $S_1$ in (\ref{Telemetry1}) and $S_2$ in (\ref{Telemetry3}) as functions of the proper time $\tau$ (or equivalently $x$ in (\ref{SeriesExpansionsDeformations})) of their emission by the probe, allow to reconstruct the deformed space-time metric order by order. To this end, we first consider the following expansion of the telemetry data
\begin{align}
&S_1=\sum_{n=1}^\infty \sa{n}\,x^n\,,&&\text{and} &&S_2=\sum_{n=1}^\infty \sd{n}\,x^n\,,
\end{align}
where we shall consider $\{\sa{n}\}$ and $\{\sd{n}\}$ as known parameters. We furthermore use the expansions (\ref{SeriesExpansionsDeformations}) of the deformation functions (\ref{DefDeformation}), where $\{\alpha_n\}$, $\{\varphi_n\}$ and $\{\psi_n\}$ are considered unknown, but satisfy the relations (\ref{RelAlpha1}) and (\ref{HigherAlphaOrders}). 

Starting with the signal $S_1$, using the form (\ref{DefDeformation}), we can write\footnote{We assume here implicitly $\Psi>0$, which has to hold at least close to the horizon for $h$ to be a smooth function with a zero at $\zh$.}
\begin{align}
&S_1=1-\sqrt{\frac{\Psi(\zp)}{\zp}}&&\text{such that} &&\Psi=\zp\,(S_1-1)^2\,.\label{Telemetry1Relation}
\end{align}
Substituting the series (\ref{SeriesExpansionsDeformations}), we obtain
\begin{align}
&\zp\,(S_1-1)^2=\left(\zh+\sum_{n=1}^\infty \alpha_n x^n\right)\left(-1+\sum_{n=1}^\infty \sa{n}\,x^n\right)^2\nonumber\\
&=\zh+(\alpha_1-2\zh\sa{1})\,x+\left(\alpha_2-2\zh\sa{2}+(\zh\sa{1}-2\alpha_1)\sa{1}\right)\,x^2\nonumber\\
&\hspace{0.5cm}+\sum_{p=3}^\infty x^p\left[ \alpha_p-2\zh\sa{p}+\sum_{\ell=1}^{p-1}(\zh\,\sa{\ell}\sa{p-\ell}-2\alpha_\ell\,\sa{p-\ell})+\sum_{n=1}^{p-2}\alpha_n\sum_{\ell=1}^{p-n-1}\sa{\ell}\sa{p-n-\ell}\right]\,,
\end{align}
such that at each order $\mathcal{O}(x^p)$, the relation (\ref{Telemetry1Relation}) amounts to the equations 
\begin{align}
&p=1:&&\psi_1=\alpha_1-2\zh\sa{1}\,,\nonumber\\
&p=2:&&\psi_2=\alpha_2-2\zh\sa{2}+(\zh\sa{1}-2\alpha_1)\sa{1}\,,\nonumber\\
&p>2:&&\psi_p=\alpha_p-2\zh\sa{p}+\sum_{\ell=1}^{p-1}(\zh\,\sa{\ell}\sa{p-\ell}-2\alpha_\ell\,\sa{p-\ell})+\sum_{n=1}^{p-2}\alpha_n\sum_{\ell=1}^{p-n-1}\sa{\ell}\sa{p-n-\ell}\,.\label{TelemetryEq1}
\end{align}
We next turn to the signal $S_2$: using (\ref{SeriesExpansionsDeformations}), equation (\ref{Telemetry3}) becomes
\begin{align}
0&=\zp^3\,S_2-\zp+\Psi\nonumber\\
&=\left(\zh+\sum_{n=1}^\infty \alpha_n x^n\right)^3\left(\sum_{k=1}^\infty \sd{k}\,x^k\right)+\sum_{p=1}^\infty (\psi_p-\alpha_p)\, x^p\,,
\end{align}
which can be further developed in the following form 
{\allowdisplaybreaks
\begin{align}
0=&\,x\left(\zh^3\,\sd{1}+\psi_1-\alpha_1\right)+x^2\left(\zh^3\sd{2}+\psi_2-\alpha_2+3\zh^2\alpha_1\sd{1}\right)\nonumber\\
&+x^3\left(\zh^3\sd{3}+\psi_3-\alpha_3+3\zh^2\left(\alpha_1\sd{2}+\alpha_2\sd{1}\right)+3\zh \alpha_1^2\sd{1}\right)\nonumber\\
&+\sum_{p=4}^\infty x^p\bigg[\zh^3\sd{p}+\psi_p-\alpha_p+3\zh^2\sum_{\ell=1}^{p-1}\alpha_\ell\sd{p-\ell}+3\zh \sum_{k=1}^{p-2}\sum_{n=1}^{p-k-1}\alpha_n\alpha_{p-k-n}\,\sd{k}\nonumber\\
&\hspace{1.8cm}+\sum_{k=1}^{p-3}\sum_{n=1}^{p-k-2}\sum_{\ell=1}^{p-k-n-1}\alpha_\ell\alpha_{p-k-n-\ell}\alpha_n\,\sd{k}\bigg]\,.
\end{align}}
Vanishing of this expression therefore yields an infinite set of equations, relating the coefficients $\{\psi_n\}$, $\{\alpha_n\}$, $\{\sd{n}\}$ and $\zh$: at each order $\mathcal{O}(x^p)$ we find 
\begin{align}
&p=1: &&0=\zh^3\,\sd{1}+\psi_1-\alpha_1\,,\nonumber\\
&p=2: &&0=\zh^3\sd{2}+\psi_2-\alpha_2+3\zh^2\alpha_1\sd{1}\,,\nonumber\\
&p=3: &&0=\zh^3\sd{3}+\psi_3-\alpha_3+3\zh^2\left(\alpha_1\sd{2}+\alpha_2\sd{1}\right)+3\zh \alpha_1^2\sd{1}\,,\nonumber\\
&p\geq 4: &&0=\zh^3\sd{p}+\psi_p-\alpha_p+3\zh^2\sum_{\ell=1}^{p-1}\alpha_\ell\sd{p-\ell}+3\zh \sum_{k=1}^{p-2}\sum_{n=1}^{p-k-1}\alpha_n\alpha_{p-k-n}\,\sd{k}\nonumber\\
& &&\hspace{0.75cm}+\sum_{k=1}^{p-3}\sum_{n=1}^{p-k-2}\sum_{\ell=1}^{p-k-n-1}\alpha_\ell\alpha_{p-k-n-\ell}\alpha_n\,\sd{k}\,.\label{TelemetryEq2}
\end{align}
The equations (\ref{RelAlpha1}), (\ref{HigherAlphaOrders}), (\ref{TelemetryEq1}) and (\ref{TelemetryEq2}) are infinitely many relations for $\{\psi_n\}$, $\{\alpha_n\}$, $\{\se{n}\}$, $\{\sd{n}\}$ and $\zh$. Indeed, from the leading equation of each set (\emph{i.e.} (\ref{RelAlpha1}) and $p=1$ in (\ref{TelemetryEq1}) and (\ref{TelemetryEq2})), we find 
\begin{align}
&\zh=\sqrt{\frac{2\sa{1}}{\sd{1}}}\,,&&\varphi_1=\alpha_1-2\zh\,\alpha_1^2\,\sa{1}\,,&&\psi_1=\alpha_1-2\zh\,\sa{1}\,.\label{Solzh}
\end{align}
Substituting these solutions into the equations at order 2 (\emph{i.e.} $p=2$ in (\ref{HigherAlphaOrders}), (\ref{TelemetryEq1}) and (\ref{TelemetryEq2})), allows to determine $\alpha_1$, such that 
\begin{align}
&\varphi_1=-\frac{\left((\sa{1})^2 \sd{1}+2\sa{1} \sd{2}-2 \sa{2} \sd{1}\right)
   \left(\sd{1} \left((\sa{1})^3-2\sa{1} \sa{2}+\sd{1}\right)+2(\sa{1})^2
   \sd{2}\right)}{2 \sqrt{2} \sqrt{\sa{1}} (\sd{1})^{7/2}}\,,\nonumber\\
&\psi_1=-\frac{9\left(\sa{1}\right)^2 \sd{1}+2\sa{1} \sd{2}-2 \sa{2} \sd{1}}{2 \sqrt{2}
   \sqrt{\sa{1}} \left(\sd{1}\right)^{3/2}}\,.\label{LeadingPsiVarphi}
\end{align}
The equations at order 2 furthermore, allow to express $\varphi_2$ and $\psi_2$ as functions of $\alpha_2$. The latter in turn is fixed (uniquely) by the equations at order 3 (\emph{i.e.} $p=3$ in (\ref{HigherAlphaOrders}), (\ref{TelemetryEq1}) and (\ref{TelemetryEq2})). In general, the set of equations (\ref{HigherAlphaOrders}), (\ref{TelemetryEq1}) and (\ref{TelemetryEq2}) at order $p>2$ is linear in $\psi_p$, $\varphi_p$ and $\alpha_{p-1}$. With (\ref{Solzh}) and (\ref{LeadingPsiVarphi}), they therefore uniquely determine the expansion coefficients $\{\varphi_n\}$ and $\{\psi_n\}$ (as well as $\zh$) in terms of the measured telemetry data $\{\sa{n}\}$ and $\{\sd{n}\}$ for all $n\in\mathbb{N}$, thus completely fixing the geometry. 
\subsection{Relation to the Spatial Radial Distance from the Horizon}\label{Sect:RelProperDistance}
In the previous Subsection, we have demonstrated how to extract the full geometry of the black hole (at least in proximity of the horizon) from the measured telemetry signals $S_{1,2}$. In doing so, we have written the metric deformation functions $\Phi$ and $\Psi$ in (\ref{DefDeformation}) as functions of the proper time measured by the probe, which constitutes a physical quantity. In previous work \cite{DelPiano:2023fiw,DelPiano:2024gvw}, the near horizon geometry has been studied in more detail, using a different -- but physically equivalent -- parametrisation in terms of the proper distance to the horizon. In order to make contact to this work (and to utilise some of the physical results obtained there), we shall provide here the relation between (expansions) of the proper distance of the probe to the horizon $\rho$ and the time for the probe to cross the horizon $x$. The proper distance to the horizon \cite{Binetti:2022xdi,DelPiano:2023fiw,DelPiano:2024gvw} and $x$ (using (\ref{DiffEqProperTime})) are computed from
\begin{align}
&\frac{d\rho}{d\zp}=\frac{1}{\sqrt{f(\zp)}}\,,&&\text{and}&&\frac{dx}{d\zp}=\sqrt{\frac{h(\zp)}{f(\zp)\left(1-h(\zp)\right)}}\,.\label{DiffEqDistance}
\end{align}
We thus find
\begin{align}
&\frac{d\rho}{dx}=\sqrt{\frac{1-h(\zp)}{h(\zp)}}=\sqrt{\frac{\Psi(x)}{\zp-\Psi(x)}}&&\text{such that} &&\left(\zp-\Psi(x)\right)\,\left(\frac{d\rho}{dx}\right)^2=\Psi(x)\,.\label{DiffEqDistTime}
\end{align}
Using the expansion (\ref{SeriesExpansionsDeformations}) as well as
\begin{align}
\rho(x)=\sum_{k=1}^\infty \rc{k}\,x^{\frac{2k-1}{2}}\,,\label{SerRhoX}
\end{align}
the coefficients $\{\rc{n}\}$ can iteratively be determined from (\ref{DiffEqDistTime})
\begin{align}
\zh+\sum_{p=1}^\infty \psi_p\,x^p=\sum_{p=0}^\infty x^p\sum_{n=1}^{p+1}(\alpha_n-\psi_n)\sum_{\ell=1}^{p-n+2}\frac{2(p-n-\ell+3)-1}{2}\,\frac{2\ell-1}{2}\,\rc{\ell}\,\rc{p-n-\ell+3}\,.\label{DistanceCompareRelation}
\end{align}
From the term of order $\mathcal{O}(x^0)$ we find the relation
\begin{align}
&\zh=\frac{(\alpha_1-\psi_1)\,\rc{1}^2}{4}&&\text{such that} &&r_1=2\sqrt{\frac{\zh}{\alpha_1-\psi_1}}\,,
\end{align}
where we have chosen the solution ensuring $\rc{1}>0$, since we assume $\rho>0$ for $x>0$. For higher $p$, we have from (\ref{DistanceCompareRelation}) the equations 
\begin{align}
&\psi_p=\sum_{n=1}^{p+1}(\alpha_n-\psi_n)\sum_{\ell=1}^{p-n+2}\frac{2(p-n-\ell+3)-1}{2}\,\frac{2\ell-1}{2}\,\rc{\ell}\,\rc{p-n-\ell+3}\,,&&\forall p\geq1\,,
\end{align} 
which for given $p$ are linear in $r_{p+1}$. These equations therefore iteratively allow to uniquely determine the coefficients $\{r_n\}$ as functions of the $\{\psi_n\}$ and $\{\alpha_n\}$.

In order to express the time for the probe to cross the horizon in terms of its radial distance to the horizon, we invert the series (\ref{SerRhoX}) by introducing
\begin{align}
x=\sum_{k=1}^\infty \tc{k}\,\rho^{2k}\,,
\end{align}
where the coefficients $\{\tc{n}\}$ can be computed through series inversion (see \emph{e.g.}~\cite{Whittaker}), and explicitly we find for the first few coefficients
\begin{align}
&\tc{1}=\frac{1}{\rc{1}^2}\,,&&\tc{2}=-\frac{2 \rc{2}}{\rc{1}^5}\,,&&\tc{3}=\frac{7 \rc{2}^2-2 \rc{1} \rc{3}}{\rc{1}^8}\,,&&\tc{4}=-\frac{2 \left(\rc{1}^2 \rc{4}-9 \rc{1} \rc{2} \rc{3}+15 \rc{2}^3\right)}{\rc{1}^{11}}\,.
\end{align}
We can thus express the coefficients $\{\xi_{2n}\}$ and $\{\theta_{2n}\}$ (for $n\geq 1$) in \cite{DelPiano:2023fiw} in the following manner\footnote{To be concrete, the functions $\Phi$ and $\Psi$ in (\ref{DefDeformation}) are expanded in powers of $\rho$ as
\begin{align}
&\Phi=\zh+\sum_{n=1}^\infty\xi_{2n}\,\rho^{2n}\,,&&\text{and} &&\Psi=\zh+\sum_{n=1}^\infty\theta_{2n}\,\rho^{2n}\,.\label{SeriesDistance}
\end{align}} 
\begin{align}
&\xi_2=t_1\,\varphi_1\,,&&\text{and} &&\xi_{2p}=\sum_{n=1}^p\sum_{{0\leq \ell_{p-1}\leq \ldots\leq\ell_1\leq n}\atop{n+\ell_1+\ldots+\ell_{p-1}=p}}\frac{\varphi_n\,n!\,\tc{1}^{n-\ell_1}\tc{2}^{\ell_1-\ell_2}\ldots \tc{p-1}^{\ell_{p-2}-\ell_{p-1}}\tc{p}^{\ell_{p-1}}}{(n-\ell_1)!(\ell_1-\ell_2)!\ldots(\ell_{p-2}-\ell_{p-1})!\ell_{p-1}!}\hspace{1cm}\forall p> 1\,.\nonumber\\
&\theta_2=t_1\,\psi_1\,,&&\text{and} &&\theta_{2p}=\sum_{n=1}^p\sum_{{0\leq \ell_{p-1}\leq \ldots\leq\ell_1\leq n}\atop{n+\ell_1+\ldots+\ell_{p-1}=p}}\frac{\psi_n\,n!\,\tc{1}^{n-\ell_1}\tc{2}^{\ell_1-\ell_2}\ldots \tc{p-1}^{\ell_{p-2}-\ell_{p-1}}\tc{p}^{\ell_{p-1}}}{(n-\ell_1)!(\ell_1-\ell_2)!\ldots(\ell_{p-2}-\ell_{p-1})!\ell_{p-1}!}\,.\hspace{1cm}\forall p> 1\,.\nonumber
\end{align}
Specifically, using the solutions (\ref{Solzh}) and (\ref{LeadingPsiVarphi}) we find for the leading coefficients
\begin{align}
\xi_2&=-\frac{\sqrt{\sa{1}} \left((\sa{1})^2 \sd{1}+2 \sa{1} \sd{2}-2 \sa{2} \sd{1}\right) \left(\sd{1} \left(\sa{1}^3-2 \sa{1} \sa{2}+\sd{1}\right)+2(\sa{1})^2 \sd{2}\right)}{4 \sqrt{2} \sd{1}^{7/2}}\,,\nonumber\\
\theta_2&=-\frac{\sqrt{\sa{1}} \left(9 \sa{1}^2 \sd{1}+2 \sa{1} \sd{2}-2 \sa{2}
   \sd{1}\right)}{4 \sqrt{2} \sd{1}^{3/2}}\,.
\end{align}
These coefficients allow us to take over some of the physical results obtained in \cite{DelPiano:2024gvw} and express them in terms of the telemetry data. For example, the Hawking temperature~\cite{Hawking:1975vcx} (which was computed via the surface gravity) of the black hole was found to be
\begin{align}
&T_H=\frac{\sqrt{1+\varpi-8\zh\,\theta_2}}{4\pi\sqrt{2}\,\zh}\,,&&\text{with} &&\varpi=\sqrt{1-16\,\zh\,\xi_2}\,,
\end{align}
which in terms of the telemetry data reads
\begin{align}
&T_H=\nonumber\\
&\frac{\sqrt{\sd{1} \left(18 (\sa{1})^3-4 \sa{1} \sa{2}+\sd{1}\right)+4 (\sa{1})^2
   \sd{2}+\left|\sd{1} \left(2 (\sa{1})^3-4 \sa{1} \sa{2}+\sd{1}\right)+4
   (\sa{1})^2 \sd{2}\right|}}{8 \pi  \sqrt{\sa{1} \sd{1}}}\,.\label{HawkingTemperature}
\end{align}
\subsection{Examples}
In order to illustrate the formalism and the results outlined in this Section, we shall discuss in the following two concrete spherically-symmetric and static black hole geometries as examples.
\subsubsection{Schwarzschild Black Hole}
\begin{figure}[htbp]
\begin{center}
\parbox{7.5cm}{\includegraphics[width=7.5cm]{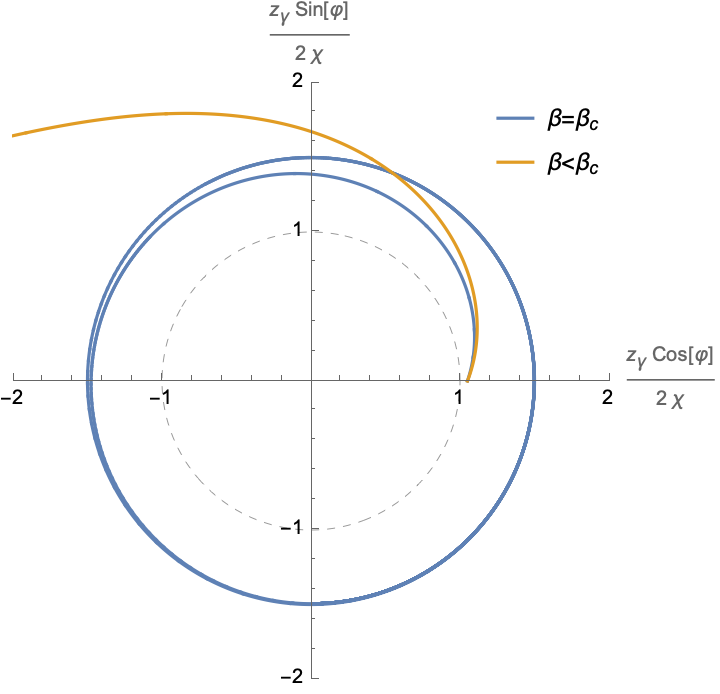}}\hspace{1cm}\parbox{7.5cm}{\includegraphics[width=7.5cm]{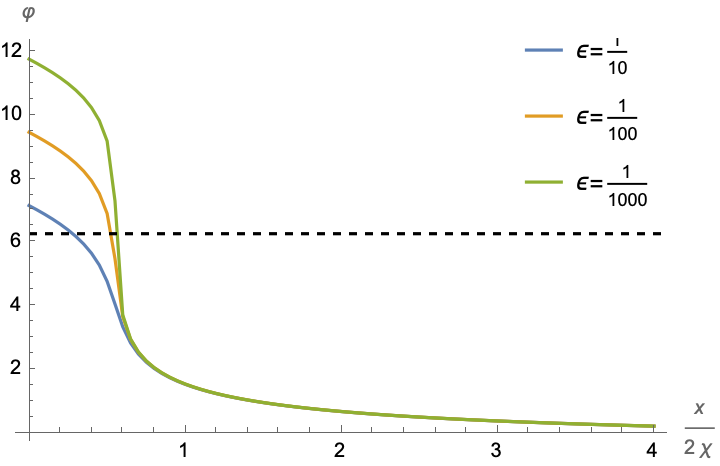}}
\end{center}
\caption{\emph{Left panel: critical versus non-critical trajectory of a photon emitted by the probe: the blue trajectory (emitted with $\beta=\beta_c$) ends on an orbit around the black hole, while the orange trajectory escapes the black hole and can be detected by the spacecraft. Both photons are emitted at a radial distance of $\zp=2.1\,\chi$, slightly outside of the horizon of the black hole (which is represented by the dashed black circle). Right panel: Angle under which photons emitted with $\beta=\beta_c-\epsilon$ are captured by the spacecraft as a function of the time when it has been emitted by the probe: $x$ is the (proper) time the probe still has to fall before reaching the horizon, when emitting the photon. The dashed line represents $2\pi$, \emph{i.e.} the case when the photon has circled the black hole once.
}}
\label{Fig:SchwarzschildPhotonTrajectories}
\end{figure}

As a first example, we consider the Schwarzschild geometry \cite{Schwarzschild:1916uq}, which is a solution of classical GR, characterised by $f(z)=h(z)=f_S(z)=1-\frac{2\chi}{z}$, where $\chi$ is the mass of the black hole. Although the deformation functions $\Phi$ and $\Psi$ in (\ref{DefDeformation}) are trivial (\emph{i.e.} constant) in this case, this example has the advantage that many of the computations of before can be performed analytically in closed form. Indeed, the proper time measured by the probe (as a function of $\zp$) and the proper time when it crosses the horizon $\tah$ are 
\begin{align}
&\tau=\frac{\sqrt{2}\left(\zs^{3/2}-\zp^{3/2}\right)}{3\sqrt{\chi}}\,,&&\text{and}&&\tah=\frac{\sqrt{2}\zs^{3/2}}{3\sqrt{\chi}}-\frac{4\chi}{3}\,.\label{Schwarzschildtau}
\end{align}
In the variable $x=\tah-\tau$, the position of the probe $\zp$ is therefore
\begin{align}
\zp=\chi\,\left(2\sqrt{2}+\frac{3x}{\sqrt{2}\chi}\right)^{2/3}\,.\label{Schwarzschildzp}
\end{align}
Figure~\ref{Fig:SchwarzschildPhotonTrajectories} compares the trajectories of photons emitted under different angles by the probe: the left panel compares the trajectory of a photon that is caught in a circular orbit to one that escapes the black hole and can be detected by the spacecraft. The right panel shows the angle under which such latter photons are detected as a function of the time when they were emitted by the probe. Furthermore, with the parametrisation (\ref{Schwarzschildtau}) and (\ref{Schwarzschildzp}), we find for the telemetry signals $S_{1,2}$ in (\ref{Telemetry1}) and (\ref{Telemetry3})

\begin{align}
&S_1(x)=1-\frac{2^{2/3}}{\left(4+\frac{3x}{\chi}\right)^{1/3}}\,,&&\text{and} &&S_2=\frac{\left(\frac{6x}{\chi} +8\right)^{2/3}-4}{\left(\frac{3x}{\chi} +4\right)^2 \chi ^2}\,.\label{SchwarzschildTelemetryExp}
\end{align}

\begin{figure}[htbp]
\begin{center}
\includegraphics[width=7.5cm]{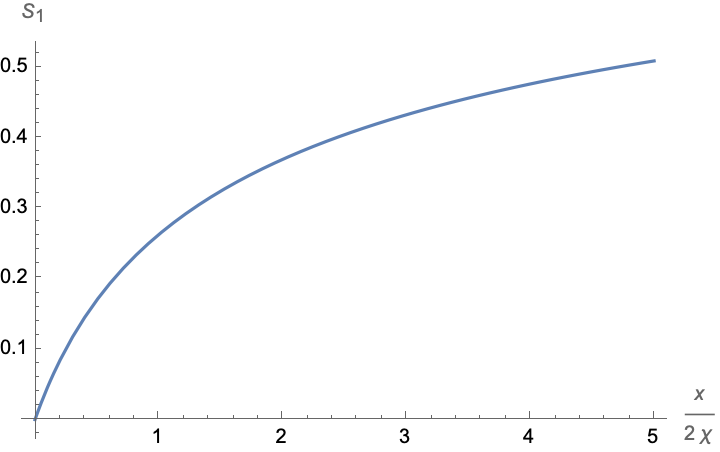}\hspace{1cm}\includegraphics[width=7.5cm]{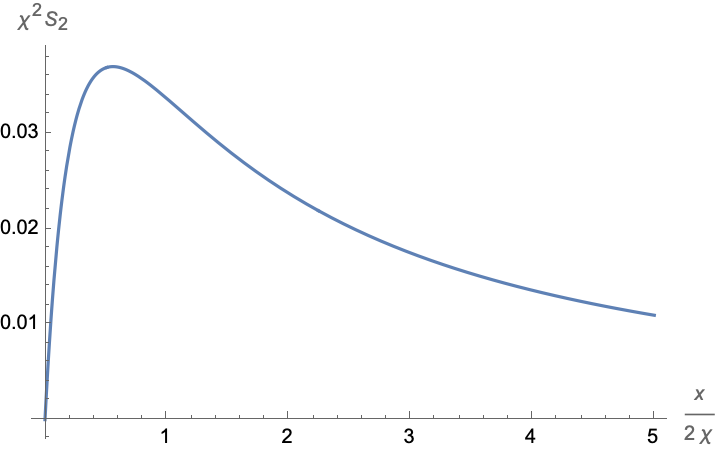}
\end{center}
\caption{\emph{The telemetries $S_1$ (left panel) and $S_2$ (right panel) as computed in (\ref{SchwarzschildTelemetryExp}), as functions of $x$ (in units of the Schwarzschild radius).
}}
\label{Fig:SchwarzschildTelemetry}
\end{figure}

\noindent
These are plotted as functions of $x$ (in units of the Schwarzschild radius $\zh=2\chi$) in Figure~\ref{Fig:SchwarzschildTelemetry}. With (\ref{SchwarzschildTelemetryExp}), the first few coefficients $\{\sa{n}\}$ and $\{\sd{n}\}$ become
\begin{align}
&\sa{1}=\frac{1}{4\chi}\,,&&\sa{2}=-\frac{1}{8\chi^2}\,,&&\sa{3}=\frac{7}{96 \chi^3}\,,&&\sa{4}=-\frac{35}{768\chi^4}\,,\nonumber\\
&\sd{1}=\frac{1}{8\chi^3}\,,&&\sd{2}=-\frac{13}{64\chi^4}\,,&&\sd{3}=\frac{23}{96\chi^5}\,,&&\sd{4}=-\frac{95}{384\chi^6}\,.\label{TelemetrySchwarzschild}
\end{align}
These coefficients lead to
\begin{align}
&\zh=2\chi\,,&&\alpha_1=1\,,&&\alpha_2=-\frac{1}{8\chi}\,,&&\alpha_3=\frac{1}{24\chi^2}\,,&&\alpha_4=-\frac{7}{384\chi^3}\,,&&\varphi_{1,2,3,4}=0=\psi_{1,2,3,4}\,,
\end{align}
which is indeed the expected (undeformed Schwarzschild) geometry. Furthermore, substituting (\ref{TelemetrySchwarzschild}) into (\ref{HawkingTemperature}), we find for the Hawking temperature $T_H=\frac{1}{8\pi\chi}$, which is indeed the expected result, thus demonstrating the consistency of our computations.
\subsubsection{Hayward Black Hole}
As an example with a non-trivial deformation function, we consider the Hayward black hole, which was introduced in \cite{Hayward_2006} and for which we shall use a similar parametrisation as in~\cite{Rezzolla:2014mua,DelPiano:2024nrl}
\begin{align}
&f(z)=h(z)=f_{\text{Hay}}(z):=1-\frac{2\chi z^2}{z^3+\frac{(2\chi)^3\gamma}{(1+\gamma)^3}}\,,&&\text{such that} &&\begin{array}{l}\Phi=\Psi=\Phi_{\text{Hay}}:=\frac{2\chi z^3}{z^3+\frac{(2\chi)^3\gamma}{(1+\gamma)^3}}\,,\\
\zh=\frac{2\chi}{1+\gamma}\,,\end{array}\label{HorizonHayward}
\end{align}
where $\gamma$ is a real, non-negative parameter, which we shall take to be small in the following. Furthermore, the position of the probe $\zp$ as a function of $x$ can be expanded as in (\ref{SeriesExpansionsDeformations}), with the following coefficients 
\begin{align}
&\alpha_1=1\,,&&\alpha_2=-\frac{1-2\gamma}{8\chi}\,,&&\alpha_3=\frac{1-7\gamma+\gamma^2}{24\chi^2}\,,&&\alpha_4=-\frac{7-105 \gamma +75 \gamma ^2-2 \gamma ^3}{384\chi^3}\,.
\end{align}
Furthermore, we find for the expansion coefficients of the telemetry signal $S_1$
\begin{align}
&\sa{1}=\frac{1-2\gamma}{4\chi}\,,\hspace{1cm}\sa{2}=-\frac{1-7\gamma+\gamma^2}{8\chi^2}\,,\hspace{1cm}\sa{3}=\frac{7-105 \gamma +75 \gamma ^2-2 \gamma ^3}{96 \chi^3}\,,\nonumber\\
&\sa{4}=-\frac{35-910 \gamma +1560 \gamma ^2-328 \gamma ^3+2 \gamma ^4}{768 \chi^4}\,,\label{TelemetryHayward1}
\end{align}
as well as $S_2$
\begin{align}
&\sd{1}=\frac{(1-2\gamma)(1+\gamma)^2}{8\chi^3}\,,\hspace{0.2cm}\sd{2}=\frac{(\gamma +1)^2 \left(8 \gamma ^2+40 \gamma -13\right)}{64 \chi ^4}\,,\hspace{0.2cm}\sd{3}=\frac{(\gamma +1)^2 \left(23-111 \gamma-3 \gamma ^2-4 \gamma ^3 \right)}{96 \chi ^5}\,,\nonumber\\
&\sd{4}=-\frac{(\gamma +1)^2 \left(380-2755 \gamma+1164 \gamma ^2-172 \gamma ^3-16 \gamma ^4\right)}{1536 \chi ^6}\,.\label{TelemetryHayward2}
\end{align}
%

\noindent
Substituting these coefficients into (\ref{Solzh}), we indeed recover $\zh$ in (\ref{HorizonHayward}). We furthermore find from (\ref{LeadingPsiVarphi}) (and the further higher solutions of (\ref{HigherAlphaOrders}), (\ref{TelemetryEq1}) and (\ref{TelemetryEq2}))
\begin{align}
&\varphi_1=\psi_1=\frac{3\gamma}{1+\gamma}\,,\hspace{1cm}\varphi_2=\psi_2=-\frac{9\gamma(3-2\gamma)}{8\chi (1+\gamma)}\,,\hspace{1cm}\varphi_3=\psi_3=\frac{9\gamma(3-6\gamma+\gamma^2)}{8\chi^2 (1+\gamma)}\,,\nonumber\\
&\varphi_4=\psi_4=-\frac{27 \gamma  \left(15-60 \gamma+33 \gamma ^2-2 \gamma ^3\right)}{128 \chi^3 (\gamma +1)}\,,\label{HaywardMetricCoefficients}
\end{align}
which indeed correctly capture the near-horizon geometry of the Hayward black hole. Finally, substituting (\ref{TelemetryHayward1}) and (\ref{TelemetryHayward2}) into (\ref{HawkingTemperature}), we obtain for the Hawking temperature
\begin{align}
T_H=\frac{1-2\gamma}{8\pi\chi}\,.
\end{align} 

\section{Relation to the Energy Momentum Tensor}
In the previous Section, we have rewritten the geometry of a spherically symmetric and static black hole (at least in a region close to the event horizon) in terms of purely physical data, concretely, the telemetric data $S_{1,2}$ received from a probe freely falling towards the black hole. Conceptually, this further strengthens the effective metric descriptions of such black holes advocated in  \cite{DelPiano:2023fiw,Binetti:2022xdi}, by providing a \emph{physical} way of determining the metric deformation functions. From a model-building perspective beyond GR, it would be interesting to also find a theoretical description of these deformations. In the following we shall therefore explain how to express the metric deformations in terms of (invariant eigenvalues of) the energy momentum tensor. This allows to interpret the geometry as solution of an effective field equation with a local energy momentum tensor, which in turn can be mimicked by suitable field configurations \cite{Boonserm:2015aqa}.\footnote{Alternatively, since we shall still parametrise the geometry in terms of the proper time of a free-falling observer, our results can also be interpreted in a similar way as in the previous Section: instead of the telemetry data $S_{1,2}$, we need to assume that the probe is capable of measuring and transmitting the energy density and pressure components of space-time to the spacecraft.}
\subsection{Energy Momentum Tensor}\label{Sect:EnergyMomentum}
We start from the general form \cite{Hawking:1973uf} (see also \cite{DAlise:2023hls}) of the energy momentum tensor $T_{\mu\nu}$ of a spherically symmetric and static black hole with the general metric (\ref{SchwarzschildMetric})
\begin{align}
&T_{\mu\nu}=\left(\begin{array}{cccc}&&&\\[-6pt]h\,\epsilon & 0 & 0 & 0 \\[3pt] 0 & \frac{p_{||}}{f} & 0 & 0 \\[3pt] 0 & 0 &  p_{\perp}\,z^2 & 0 \\[3pt] 0 & 0 & 0 & p_\perp\,z^2\,\sin^2\theta\end{array}\right)\,,&&\text{with} &&\begin{array}{l} \epsilon=\frac{1-f-zf'}{8\pi z^2}\,, \\[6pt] p_{||}=\frac{f-1+zf \frac{h'}{h}}{8\pi z^2}\,,\\[10pt] p_{\perp}=\frac{z}{2}p'_{||}+\frac{zh'}{4h}(\epsilon+p_{||})+p_{||}\,.\end{array}\label{DefEnergyMomentum}
\end{align}
Here $\epsilon$, $p_{||}$ and $p_{\perp}$ are the Lorentz invariant eigenvalues of of ${T^\mu}_\nu$, which are called the energy density, the parallel and perpendicular pressure component respectively. It is crucial to realise, since they are constructed from only 2 independent functions (namely $h$ and $f$), these three eigenvalues are functionally not completely independent. In the following, we shall rewrite $\epsilon$, $p_{||}$ and $p_{\perp}$ in terms of the proper time $x$ of a free falling observer to cross the horizon, as discussed in Section~\ref{Sect:FreeFallingObserver}:
\begin{itemize}
\item {\bf energy density:} For $\epsilon$ in (\ref{DefEnergyMomentum}) we introduce the following series expansion in $x=\tah-\tau$
\begin{align}
\epsilon=\frac{d\Phi}{d\tau}\,\frac{\tau'(z)}{8\pi z^2}=\sum_{n=0}^\infty \epsilon_n x^n\,. 
\end{align}
Using (\ref{SeriesExpansionsDeformations}) as well as the expansions of $1/z^2$ and $\tau'(z)=\frac{d\tau}{dz}$ worked out in Appendix~\ref{App:PowerInversion}, we can iteratively determine the coefficients $\epsilon_p$
\begin{align}
&\epsilon_0=-\frac{\varphi_1 \mathfrak{p}_{0}}{8\pi\zh^2}\,,
&&\epsilon_p=-\frac{1}{8\pi}\left[\frac{1}{\zh^2}\sum_{n=1}^{p+1}n\varphi_n\,\mathfrak{p}_{p-n+1}+\sum_{m=1}^{p}\mathfrak{q}_m^{(2)}\sum_{n=1}^{p-m+1}n\varphi_n\mathfrak{p}_{p-m-n+1}\right]\,,\hspace{0.5cm}\forall p>0\,,\label{EpsilonFullExpansion}
\end{align}
where $\{\mathfrak{p}\}$ are iteratively defined in (\ref{ExpansionCoDer}) and $\{\mathfrak{q}_n^{(2)}\}$ in (\ref{ExpansionCoPowers}) as functions of the $\{\alpha_n\}$, which (along with $\{\varphi_n\}$) appear in (\ref{SeriesExpansionsDeformations}). For later convenience, we provide explicitly the first two of these coefficients
\begin{align}
&\epsilon_0=\frac{\varphi_1}{8\pi \zh^2\alpha_1}\,,&&\text{and} &&\epsilon_1=\frac{\alpha_1 \varphi_2 \zh-\alpha_1^2 \varphi_1 -\alpha_2\varphi_1 \zh }{4 \pi 
   \alpha_1^2 \zh ^3}\,.\label{Epsilon0Cond}
\end{align}
Notice, while $\epsilon_0$ is a non-linear function of $\varphi_1$ and $\psi_1$ (in particular, since $\alpha_1$ is through (\ref{RelAlpha1}) implicitly also a function of these coefficients), we obtain a linear relation between $\varphi_2$ and $\epsilon_1$. In fact, all equations for $p\geq 0$ in (\ref{EpsilonFullExpansion}) are linear in $\epsilon_p$ and $\varphi_{p+1}$.

\item {\bf parallel pressure component:} We next consider $p_{||}$ in (\ref{DefEnergyMomentum}), for which we introduce the following series expansion
\begin{align}
p_{||}=\frac{\Psi-\Phi-(z-\Phi)\frac{d\Psi}{d\tau}\,\tau'(z)}{8\pi z^2(z-\Psi)}=\sum_{n=0}^\infty \ppar{n} x^n\,.\label{ParaCond}
\end{align}
The coefficients $\ppar{p}$ for all $p\geq 0$ are iteratively given by
\begin{align}
\ppar{p}=\frac{1}{8\pi}\sum_{m=-1}^{p-1}\beta_m\left[\psi_{p-m}-\varphi_{p-m}+\sum_{n=0}^{p-m-1}\sum_{\ell=1}^{p-m-n}\ell\,(\alpha_{p-m-n-\ell+1}-\varphi_{p-m-n-\ell+1})\psi_\ell \mathfrak{p}_n\right]\,,\label{EqPara}
\end{align}
where as a shorthand we have introduced $\{\beta_n\}$ as the expansion coefficients of
\begin{align}
&\frac{1}{z^2(z-\Psi)}=\sum_{p=-1}^\infty \beta_p\,x^p\,,&&\text{with} &&\begin{array}{l} \beta_{-1}=\frac{1}{\zh^2(\alpha_1-\psi_1)} \,, \\[8pt] \beta_0=-\frac{2\alpha_1(\alpha_1-\psi_1)+\zh (\alpha_2-\psi_2)}{\zh^3(\alpha_1-\psi_1)^2}\,,\end{array}
\end{align}
and all $\beta_{p\geq 2}$ are iteratively fixed through
\begin{align}
&-\zh^2(\alpha_1-\psi_1)\beta_{p}=\zh\,\beta_{p-1}\left(\zh(\alpha_2-\psi_2)+2\alpha_1(\alpha_1-\psi_1)\right)\nonumber\\
&\hspace{0.5cm}+\sum_{n=3}^{p+2}\beta_{p+1-n}\left[\zh^2(\alpha_n-\psi_n)+2\zh\sum_{\ell=1}^{n-1}\alpha_{n-\ell}(\alpha_\ell-\psi_\ell)+\sum_{\ell=1}^{n-2}\sum_{k=1}^{n-\ell-1}\alpha_k\alpha_{n-\ell-k}(\alpha_\ell-\psi_\ell)\right]\,.
\end{align}
Notice that $\ppar{0}=-\epsilon_0$, while we have for $\ppar{1}$
\begin{align}
\ppar{1}=\frac{2 \alpha_1^3 \varphi_1-\alpha_1^2 (2 \varphi_1 \psi_1+\zh  (\varphi_2+\psi_2))+\alpha_1 \zh  (\alpha_2 (\varphi_1+\psi_1)+\varphi_1 \psi_2+\varphi_2
   \psi_1)-2 \alpha_2 \varphi_1 \psi_1 \zh }{8 \pi  \alpha_1^2 \zh ^3 (\alpha_1-\psi_1)}\,,\label{ParCon1}
\end{align}
which is linear in $\varphi_2$, $\psi_2$ and $\alpha_2$.

\item {\bf perpendicular pressure component:} Finally we introduce the series expansion of $p_{\perp}$
\begin{align}
p_{\perp}=\frac{z}{2}\,\frac{dp_{||}}{dz}+\frac{z h'}{4h}(\epsilon+p_{||})+p_{||}=\frac{z}{2}\,\frac{dp_{||}}{d\tau}\,\tau'(z)+\frac{\Psi-z\frac{d\Psi}{d\tau}\,\tau'(z)}{4(z-\Psi)}(\epsilon+p_{||})+p_{||}=\sum_{n=0}^\infty \pperp{n} x^n\,.\label{PerpCond}
\end{align}
The coefficients $\pperp{p}$ are iteratively given by
\begin{align}
\pperp{0}&=\frac{3 \ppar{1} \zh+\zh \epsilon_1-4 \alpha_1 \epsilon_0}{4 \alpha_1}\,,\nonumber\\
\pperp{p}&=-\frac{\zh}{2}\sum_{\ell=1}^{p+1}\ell \ppar{\ell}\mathfrak{p}_{p-\ell+1}-\frac{1}{2}\sum_{n=1}^p\alpha_n\sum_{\ell=1}^{p-n+1}\ell\,\ppar{\ell}\mathfrak{p}_{p-n-\ell+1}+\ppar{p}\nonumber\\
&\hspace{2cm}+\frac{1}{4}\sum_{k=-1}^{p-1}(\epsilon_{p-k}+\ppar{p-k})\sum_{n=0}^{k+1}\delta_{n}\gamma_{k-n}\,,\hspace{1cm}\forall p\geq 1\,,\label{PerpCon0}
\end{align}
where as a shorthand notation we have introduced the following expansion coefficients
\begin{align}
&\frac{1}{z-\Psi}=\sum_{n=-1}^\infty \gamma_n x^n&&\text{with} &&\left\{\begin{array}{ll}\gamma_{-1}=\frac{1}{\alpha_1-\psi_1}\,, &\\[4pt]\gamma_{p}=-\frac{1}{\alpha_1-\psi_1}\sum_{\ell=-1}^{p-1}\gamma_{\ell}(\alpha_{p-\ell+1}-\psi_{p-\ell+1})\,, &\forall p\geq 0\,,\end{array}\right.\nonumber\\[8pt]
&\Psi-z\frac{d\psi}{d\tau}\,\tau'(z)=\sum_{n=0}^\infty\delta_n x^n\,,&&\text{with} &&\left\{\begin{array}{ll}\delta_0=\frac{\zh(\alpha_1-\psi_1)}{\alpha_1}\,,&\\[4pt]\delta_p=\psi_p+\zh\sum_{\ell=1}^{p+1} \ell \psi_\ell \mathfrak{p}_{p-\ell+1}\\[8pt] \hspace{1cm}+\sum_{n=1}^p\alpha_n\sum_{\ell=1}^{p-n+1}\ell\psi_\ell \mathfrak{p}_{p-n-\ell+1}\,, &\forall p\geq 0\,.\end{array}\right.\nonumber
\end{align}
For later convenience, we specify $\pperp{1}$
\begin{align}
\pperp{1}=\frac{1}{4 \alpha_1^2 (\alpha_1-\psi_1)}\bigg[&\ppar{1} \left(6 \alpha_1^3-6 \alpha_1^2 \psi_1-\alpha_1\zh (5 \alpha_2+\psi_2)+6
   \alpha_2 \psi_1\zh\right)+5 \alpha_1 \ppar{2}\zh (\alpha_1-\psi_1)\nonumber\\
   &+\zh
   \left(\alpha_1^2 \epsilon_2-\alpha_1 (\alpha_2 \epsilon_1+\psi_2 \epsilon_1+\psi_1 \epsilon_2)+2 \alpha_2 \psi_1 \epsilon_1\right)\bigg]\,.\label{PerpCon1}
\end{align}
\end{itemize}
\subsection{Iterative Solution}\label{Sect:EMTiterativeSol}
Before moving on to discuss potential directions towards an effective field equation for the metric that is encoded in the coefficients $\{\epsilon_n,\ppar{n},\pperp{n}\}$, we remark that they can in fact (also) be used to uniquely fix the metric deformations. To leading order, the equations for $\epsilon_0$ in (\ref{Epsilon0Cond}), for $\pperp{0}$ in (\ref{PerpCon0}) as well as (\ref{RelAlpha1}) (and using that $\epsilon_0=-\ppar{0}$), can be inverted to give 
\begin{align}
&\varphi_1=\frac{2\pi \epsilon_0\,\zh^3 (3\ppar{1}+\epsilon_1)}{\pperp{0}+\epsilon_0}\,,&&\psi_1=\frac{4 (\pperp{0}+\epsilon_0) \left(8 \pi  \zh^2 \epsilon_0-1\right)}{\zh (3\ppar{1}+\epsilon_1)}+\frac{\zh (3\ppar{1} +\epsilon_1)}{4 (\pperp{0}+\epsilon_0)}\,,&&\alpha_1=\frac{\zh (3\ppar{1}+\epsilon_1)}{\pperp{0}+\epsilon_0}\,,\label{SolutionLevel1}
\end{align}
thus fixing the leading coefficients of the metric deformations (\ref{DefDeformation}) fully in terms of the expansion coefficients of the energy momentum tensor as well as $\zh$. 
At the next order, the equations (\ref{Epsilon0Cond}), (\ref{ParCon1}) and (\ref{PerpCon1}), as well as (\ref{HigherAlphaOrders}) for $p=2$ not only fix the coefficients $\varphi_2$, $\psi_2$ and $\alpha_2$, but also impose a consistency relation among $\zh$ and $\{\epsilon_0\,,\epsilon_1\,,\epsilon_2\,,\ppar{1}\,,\ppar{2}\,,\pperp{0}\,,\pperp{1}\,,\pperp{2}\}$, which can be solved in the form $\zh=\sqrt{\frac{N}{8\pi D}}$, with the shorthand notation
{\allowdisplaybreaks
\begin{align}
&N=-27{\ppar{1}}^3+18{\ppar{1}}^2 ({\pperp{1}}-\epsilon_1)-(3\ppar{1}+\epsilon_1) \bigg[81{\ppar{1}}^4+54{\ppar{1}}^3 (\epsilon_1-2
   {\pperp{1}})+{\ppar{1}}^2 \bigg(9 \left(4 {\pperp{1}}^2-8 {\pperp{1}} \epsilon_1+\epsilon_1^2\right)\nonumber\\
   &\hspace{1cm}-4
   ({\pperp{0}}+\epsilon_0) (-45\ppar{2}+40 \pi  ({\pperp{0}}+\epsilon_0) (3 {\pperp{0}}+2 \epsilon_0)-9
   \epsilon_2)\bigg)-12\ppar{1} \bigg(5\ppar{2} ({\pperp{0}}+\epsilon_0) (2 {\pperp{1}}-\epsilon_1)\nonumber\\
   &\hspace{1cm}-(2
   {\pperp{1}}-\epsilon_1) ({\pperp{1}} \epsilon_1-\epsilon_2 ({\pperp{0}}+\epsilon_0))+16 \pi  \epsilon_1 (3 {\pperp{0}}+2 \epsilon_0) ({\pperp{0}}+\epsilon_0)^2\bigg)\nonumber\\
   &\hspace{1cm}+4 \bigg(25{\ppar{2}}^2 ({\pperp{0}}+\epsilon_0)^2+10\ppar{2} ({\pperp{0}}+\epsilon_0) (\epsilon_2 ({\pperp{0}}+\epsilon_0)-{\pperp{1}} \epsilon_1)+({\pperp{1}} \epsilon_1-\epsilon_2 ({\pperp{0}}+\epsilon_0))^2\nonumber\\
   &\hspace{1cm}-8 \pi  \epsilon_1^2 (3 {\pperp{0}}+2
   \epsilon_0) ({\pperp{0}}+\epsilon_0)^2\bigg)\bigg]^{1/2}+{\ppar{1}} (2 ({\pperp{0}}+\epsilon_0) (-15\ppar{2}+80 \pi 
   \epsilon_0 ({\pperp{0}}+\epsilon_0)-3 \epsilon_2)\nonumber\\
   &\hspace{1cm}+3 \epsilon_1 (4 {\pperp{1}}-\epsilon_1))+2
   \epsilon_1 (({\pperp{0}}+\epsilon_0) (-5\ppar{2}+16 \pi  \epsilon_0 ({\pperp{0}}+\epsilon_0)-\epsilon_2)+{\pperp{1}} \epsilon_1)\,,\nonumber\\
&D=9{\ppar{1}}^3 (3 {\pperp{0}}-4 \epsilon_0)+{\ppar{1}}^2 (45 {\pperp{0}} \epsilon_1+36 {\pperp{1}} \epsilon_0-6
   \epsilon_0 \epsilon_1)+{\ppar{1}} \bigg(-60\ppar{2} \epsilon_0 ({\pperp{0}}\nonumber\\
   &\hspace{1cm}+\epsilon_0)+\epsilon_1 (21 {\pperp{0}} \epsilon_1+8 \epsilon_0 (3 {\pperp{1}}+\epsilon_1))+160 \pi  \epsilon_0^2
   ({\pperp{0}}+\epsilon_0)^2-12 \epsilon_0 \epsilon_2 ({\pperp{0}}+\epsilon_0)\bigg)\nonumber\\
   &\hspace{1cm}+\epsilon_1
   \left(-20\ppar{2} \epsilon_0 ({\pperp{0}}+\epsilon_0)+\epsilon_1 (3 {\pperp{0}} \epsilon_1+4 {\pperp{1}}
   \epsilon_0+2 \epsilon_0 \epsilon_1)+32 \pi  \epsilon_0^2 ({\pperp{0}}+\epsilon_0)^2-4 \epsilon_0 \epsilon_2 ({\pperp{0}}+\epsilon_0)\right)\,.
\end{align}}
 Together with (\ref{SolutionLevel1}) we therefore find a solution for $(\varphi_1\,,\varphi_2\,,\psi_1\,,\psi_2)$ entirely in terms of $\{\epsilon_0\,,\epsilon_1\,,\epsilon_2\,,\ppar{1}\,,\ppar{2}\,,\pperp{0}\,,\pperp{1}\,,\pperp{2}\}$, which also fixes the position $\zh$ of the horizon.
 
All further equations (\ref{EpsilonFullExpansion}), (\ref{EqPara}) and (\ref{PerpCon0}) for $p> 1$ are linear relations that (uniquely) fix $\psi_p$ and $\varphi_p$ respectively while still imposing further consistency conditions among the eigenvalues of the energy momentum tensor.\footnote{If these conditions are not satisfied, the black hole is in fact not spherically symmetric and static, but is instead parametrised by additional degrees of freedom.} Thus, in total the latter completely determine the metric deformations: in other words, if the probe considered in the previous Section was capable of measuring (and transmitting) the eigenvalues of the energy momentum tensor $\epsilon$, $p_{||}$ and $p_\perp$ as functions of the proper time instead of the telemetry data $S_{1,2}$, we would equally be able to reconstruct the full black hole geometry from it (at least close to the horizon).

\subsection{Towards an Effective Field Description}
\subsubsection{Energy Momentum Tensor and Minimal Field Description}
In addition to providing a viable alternative to the telemetry data of the previous Section, there is further interest in the equations (\ref{EpsilonFullExpansion}), (\ref{EqPara}) and (\ref{PerpCon0})  that establish a relation between the eigenvalues of the energy momentum tensor and the black hole geometry. Indeed, in GR the energy momentum tensor enters as an inhomogeneity of the Einstein equation
\begin{align}
R_{\mu\nu}-\frac{1}{2}\,R\,g_{\mu\nu}=8\pi\,T_{\mu\nu}\,,\label{EinsteinEquation}
\end{align}
where $R_{\mu\nu}$ and $R$ are the Ricci tensor and Ricci scalar respectively, which are completely determined by the geometry (\emph{i.e.} in our case the metric functions $f$ and $h$ as well as their derivatives). While the field equation for the geometry in a theory beyond GR is unknown (if it is indeed still a geometric description), we can read (\ref{EinsteinEquation}) as an effective equation for the deformed metric (\ref{SchwarzschildMetric}), for a suitable $T_{\mu\nu}$. While it is not possible to link the latter unambiguously to a physical system, a minimal configuration of physical fields representing a general (physically reasonable energy momentum tensor), was presented in \cite{Boonserm:2015aqa}.

Indeed, following the classification of \cite{Hawking:1973uf}, the energy momentum tensor (\ref{DefEnergyMomentum}) is of type I. Abstractly, it represents an anisotropic fluid, which can be represented through different field configurations. In \cite{Boonserm:2015aqa}, for a static and spherically symmetric case, a minimal such realisation was determined in the form of a massless scalar field (which we shall denote $\phi$), an electric-like field (which we shall denote $E$) and a perfect fluid that is characterised by the energy momentum tensor
\begin{align}
&{(T_f)^\mu}_\nu=\left(\begin{linesmall}{@{}cccc@{}}&&&\\[-6pt]\rho_f & 0 & 0 & 0 \\[3pt] 0 & p_f & 0 & 0 \\[3pt] 0 & 0 & p_f & 0 \\[3pt] 0 & 0 & 0 & p_f\end{linesmall}\right)\,,&&\text{with} &&p_f=\omega(\rho_f)\,,
\end{align}
where $\rho_f$ is the energy density and $p_f$ the pressure component, which are related through the equation of state $\omega$. Concretely, in \cite{Boonserm:2015aqa} the following expression of the eigenvalues $(\epsilon,p_{||},p_\perp)$ of (\ref{DefEnergyMomentum}) were given (see also \cite{DAlise:2023hls})
\begin{align}
&\epsilon=\rho_f+\frac{1}{2}\,E^2+\frac{1}{2}\,(\vec{\nabla}\phi)^2\,,&&p_{||}=p_f-\frac{1}{2}\,E^2+\frac{1}{2}\,(\vec{\nabla}\phi)^2\,,&&p_{\perp}=p_f+\frac{1}{2}\,E^2-\frac{1}{2}\,(\vec{\nabla}\phi)^2\,,
\end{align}
which has the following minimal inversion
\begin{align}
&p_f=\frac{p_{||}+p_\perp}{2}\,,&&\rho_f=\epsilon-\frac{|p_{||}+p_\perp|}{2}\,,&&(\vec{\nabla}\phi)^2=\text{max}\left[p_{||}-p_\perp,0\right]\,,&&E^2=\text{max}\left[p_\perp-p_{||},0\right]\,.\label{PhysicalSystem}
\end{align}
The iterative solutions (\ref{EpsilonFullExpansion}), (\ref{EqPara}) and (\ref{PerpCon0}) along with (\ref{SolutionLevel1}) therefore give explicit expressions for the eigenvalues $(\epsilon,p_{||},p_\perp)$ in terms of the (expansion coefficients to the) deformation functions $\Psi$ and $\Phi$. Along with (\ref{PhysicalSystem}), they therefore characterise the physical system, whose energy momentum tensor enters into the effective equation (\ref{EinsteinEquation}).
\subsubsection{Approximation}
The relation (\ref{PhysicalSystem}) links the physical system composed of a perfect fluid, an electric-like field and a scalar field, to the metric deformation functions at the level of the expansion coefficients for small $x$. In this Subsection, we shall make this relation more concrete by assuming two simplifications
\begin{itemize}
\item[\emph{(i)}] \emph{small deformations:} We shall assume that the corrections encoded in the deformation functions $\Psi$ and $\Phi$, relative to the Schwarzschild geometry, are small, which is in particular expected, if they are due to quantum effects. More concretely, we shall assume that these corrections can be encoded in a small parameter, with respect to which $\Psi$ and $\Phi$ are analytic functions. In order to define such a parameter, we shall follow a similar discussion as in \cite{DelPiano:2024nrl}: we first parametrise the deformation of the horizon position and introduce the following re-scaled deformation parameters $\{v_n,u_n\}_{n\in\mathbb{N}}$
\begin{align}
&\zh=2\chi(1+\mathfrak{c})\,,&&\varphi_n=:\frac{v_n}{(2\chi)^{n-1}}\,,&&\psi_n=:\frac{u_n}{(2\chi)^{n-1}}\,,&&\forall n\in\mathbb{N}\,.\label{MassRescaling}
\end{align}
We shall then assume that $\mathfrak{c}$ as well as $\{v_n,u_n\}_{n\in\mathbb{N}}$ are 'small' deformations, by considering
\begin{align}
&\mathfrak{c}\ll 1\,,&&\text{and} &&\mathfrak{v}_n:=\frac{v_n}{\mathfrak{c}}\gg \mathfrak{c}\,,&&\text{and} &&\mathfrak{u}_n:=\frac{u_n}{\mathfrak{c}}\gg \mathfrak{c}\,,&&\forall n\in\mathbb{N}\,.\label{ScalingApprox}
\end{align}
This hierarchy is realised in a number of examples of spherically symmetric and static (quantum) black holes discussed in the literature~\cite{bardeen1968,Dymnikova:1992ux,Dymnikova:2004qg,Hayward:2005gi,Frolov:2016pav,Simpson:2018tsi,Simpson:2019mud,Garcia95}. This approximation allows us for example to express the coefficients $\{\alpha_n\}_{n\in\mathbb{N}}$ to leading order\footnote{Following the notation in \cite{DelPiano:2024nrl} we shall use $\sim$ for relations that hold to leading order in $\mathfrak{c}$ but receive corrections at higher orders $\mathfrak{o}(\mathfrak{c})$.} in $\mathfrak{c}$ as linear combinations of $\{\mathfrak{v}_n,\mathfrak{u}_n\}_{n\in\mathbb{N}}$
\begin{align}
&\alpha_n\sim\frac{3^n\,\Gamma(5/3)}{2^n\,(2\chi)^{n-1} \Gamma(n+1)\Gamma(5/3-n)}+\frac{\mathfrak{c}}{(2\chi)^{n-1}}\left[\mathcal{A}^{(\mathfrak{c})}_n+\sum_{m=1}^\infty\left(\mathcal{A}^{(v)}_{nm}\,\mathfrak{v}_m+\mathcal{A}^{(u)}_{nm}\,\mathfrak{u}_m\right)\right]\,,\label{LeadExpansionAlpha}
\end{align}
where we explicitly find
\begin{align}
&\mathcal{A}^{(\mathfrak{c})}=\left(\begin{linesmall}{@{}c@{}} \\[-6pt] 0 \\[6pt] \frac{1}{4}\\[6pt] -\frac{1}{3}\\[6pt]\frac{7}{16}\\ \vdots  \\[2pt]
\end{linesmall}\right)\,.
&&\mathcal{A}^{(v)}=\left(\begin{linesmall}{@{}ccccc@{}}&&&&\\[-6pt] -\frac{1}{2} & 0 & 0 & 0 & \cdots \\[4pt]
 \frac{3}{16} & -\frac{1}{4} & 0 & 0 & \cdots \\[4pt]
 -\frac{29}{144} & \frac{1}{12} & -\frac{1}{6} & 0 & \cdots \\[4pt]
 \frac{1115}{4608} & -\frac{35}{384} & \frac{5}{96} & -\frac{1}{8} &\cdots  \\
 \vdots & \vdots & \vdots & \vdots & \ddots \\[2pt]
\end{linesmall}\right)\,,&&
\mathcal{A}^{(u)}=\left(\begin{linesmall}{@{}ccccc@{}}&&&&\\[-6pt] 
\frac{1}{2} & 0 & 0 & 0 & \cdots \\[4pt]
 \frac{1}{16} & \frac{1}{4} & 0 & 0 & \cdots \\[4pt]
 \frac{11}{144} & \frac{1}{12} & \frac{1}{6} & 0 & \cdots \\[4pt]
 -\frac{539}{4608} & \frac{1}{128} & \frac{7}{96} & \frac{1}{8} & \cdots \\
 \vdots & \vdots & \vdots & \vdots & \ddots \\[2pt]
\end{linesmall}\right)\,,\nonumber
\end{align}
Similarly, we can write to leading order in $\mathfrak{c}$ for the coefficients $\{\epsilon_n,\ppar{n},\pperp{n}\}_{n\in\mathbb{N}}$
\begin{align}
&\epsilon_n\sim\frac{\mathfrak{c}}{8\pi (2\chi)^{n+2}}\sum_{m=1}^\infty \mathcal{E}_{nm}\,\mathfrak{v}_m\,,\nonumber\\
&\ppar{n}\sim\frac{\mathfrak{c}}{8\pi (2\chi)^{n+2}}\sum_{m=1}^\infty \left[\mathcal{P}^{(v)}_{nm}\,\mathfrak{v}_m+\mathcal{P}^{(u)}_{nm}\,\mathfrak{u}_m\right]\,,\nonumber\\
&\pperp{n}\sim\frac{\mathfrak{c}}{8\pi (2\chi)^{n+2}}\sum_{m=1}^\infty \left[\mathcal{Q}^{(v)}_{nm}\,\mathfrak{v}_m+\mathcal{Q}^{(u)}_{nm}\,\mathfrak{u}_m\right]\,,\label{LeadExpansionEPP}
\end{align}
with the coefficients taking the form
{\allowdisplaybreaks
\begin{align}
&\mathcal{E}=\left(\begin{linesmall}{@{}ccccc@{}}&&&&\\[-6pt] 
 1 & 0 & 0 & 0 & \cdots \\[4pt]
 -\frac{3}{2} & 2 & 0 & 0 & \cdots \\[4pt]
 \frac{9}{4} & -3 & 3 & 0 & \cdots \\[4pt]
 -\frac{27}{8} & \frac{9}{2} & -\frac{9}{2} & 4 & \cdots \\
 \vdots & \vdots & \vdots & \vdots & \ddots \\[2pt]
\end{linesmall}\right)\,,\hspace{0.2cm}
\mathcal{P}^{(v)}=\left(\begin{linesmall}{@{}ccccc@{}}&&&&\\[-6pt] 
 -1 & 0 & 0 & 0 & \cdots \\[4pt]
 \frac{7}{4} & -1 & 0 & 0 & \cdots \\[4pt]
 -\frac{139}{48} & \frac{7}{4} & -1 & 0 & \cdots \\[4pt]
 \frac{299}{64} & -\frac{139}{48} & \frac{7}{4} & -1 & \cdots \\
 \vdots & \vdots & \vdots & \vdots & \ddots \\[2pt]
\end{linesmall}\right)\,,\hspace{0.2cm}
\mathcal{P}^{(u)}=\left(\begin{linesmall}{@{}ccccc@{}}&&&&\\[-6pt] 
 0 & 0 & 0 & 0 & \cdots \\[4pt]
 -\frac{1}{4} & -1 & 0 & 0 & \cdots \\[4pt]
 \frac{31}{48} & \frac{5}{4} & -2 & 0 & \cdots \\[4pt]
 -\frac{83}{64} & -\frac{77}{48} & \frac{11}{4} & -3  & \cdots \\
 \vdots & \vdots & \vdots & \vdots & \ddots \\[2pt]
\end{linesmall}\right)\,,\nonumber\\
&\mathcal{Q}^{(v)}=\left(\begin{linesmall}{@{}ccccc@{}}&&&&\\[-6pt] 
  -\frac{1}{16} & -\frac{1}{4} & 0 & 0 & \cdots \\[4pt]
 \frac{1}{48} & -\frac{1}{4} & -\frac{1}{2} & 0 & \cdots \\[4pt]
 \frac{35}{768} & \frac{21}{64} & -\frac{7}{16} & -\frac{3}{4} & \cdots \\[4pt]
 -\frac{737}{4608} & -\frac{167}{384} & \frac{61}{96} & -\frac{5}{8} & \cdots \\
 \vdots & \vdots & \vdots & \vdots & \ddots \\[2pt]
\end{linesmall}\right)\,,
\hspace{1cm}\mathcal{Q}^{(u)}=\left(\begin{linesmall}{@{}ccccc@{}}&&&&\\[-6pt] 
-\frac{3}{16} & -\frac{3}{4} & 0 & 0 & \cdots \\[4pt]
 \frac{17}{48} & -\frac{1}{4} & -\frac{5}{2} & 0 & \cdots \\[4pt]
 -\frac{467}{768} & \frac{27}{64} & -\frac{5}{16} & -\frac{21}{4} & \cdots \\[4pt]
 \frac{4625}{4608} & -\frac{265}{384} & \frac{47}{96} & -\frac{3}{8}  & \cdots \\
 \vdots & \vdots & \vdots & \vdots & \ddots \\[2pt]
\end{linesmall}\right)\,,
\end{align}
}
\item[\emph{(ii)}] $f=h$: we consider the simplifying case $\Psi=\Phi$. Examples of (quantum) deformed black hole geometries of this type have been previously studied in the literature in \emph{e.g.}~\cite{bardeen1968,Dymnikova:1992ux,Dymnikova:2004qg,Hayward:2005gi,Frolov:2016pav,Simpson:2018tsi,Simpson:2019mud}. In this way, we also impose $\mathfrak{v}_n=\mathfrak{u}_n$ $\forall n\in\mathbb{N}$, which simplifies the expressions (\ref{LeadExpansionAlpha}) and (\ref{LeadExpansionEPP}) for the coefficients $\{\alpha_n,\epsilon_n,\ppar{n},\pperp{n}\}_{n\in\mathbb{N}}$ at leading order in $\mathfrak{c}$. In fact, by studying these expressions up to order $n=12$, \emph{i.e.} the matrices $\mathcal{A}^{(\mathfrak{c},v,u)}$, $\mathcal{E}$, $\mathcal{P}^{(v,u)}$ and $\mathcal{Q}^{(v,u)}$, we observe that they follow a general pattern which fits with
\begin{align}
&\sum_{n=0}^\infty \frac{\mathcal{A}_n^{(\mathfrak{c})} x^n}{(2\chi)^{n-1}}=\frac{(x+4\chi)}{\left(8+\frac{6x}{\chi}\right)^{1/3}}-2\chi\,,&\sum_{n=0}^\infty \frac{(\mathcal{A}_{nm}^{(v)}+\mathcal{A}_{nm}^{(v)}) x^n}{(2\chi)^{n-1}}=\frac{x^{m+1}}{(2\chi)^m(m+1)\left(8+\frac{6x}{\chi}\right)^{1/3}}\,,\label{ResumAlpha}
\end{align}
as well as
\begin{align}
&\sum_{n=0}^\infty\frac{\mathcal{E}_{nm} x^n}{8\pi(2\chi)^{n+2}}=-\sum_{n=0}^\infty\frac{(\mathcal{P}^{(v)}_{nm}+\mathcal{P}^{(u)}) x^n}{8\pi(2\chi)^{n+2}}=\frac{mx^{m-1}}{4\pi(3x+4\chi)(2\chi)^m}\,,\label{ResumEpsilon}\\
&\sum_{n=0}^\infty\frac{(\mathcal{Q}^{(v)}_{nm}+\mathcal{Q}^{(u)}) x^n}{8\pi(2\chi)^{n+2}}=-\frac{m(m-1) x^{m-2}}{16\pi (2\chi)^m}-\frac{m x^{m-1}}{16 \pi(3x+4\chi)(2\chi)^m}\,.
\end{align}
\end{itemize}
These simplifying assumptions allow us to find a much more compact expression for $\zp$. Indeed, upon using (for $x<\frac{4\chi}{3}$)
\begin{align}
\sum_{n=1}^\infty \frac{3^n\,x^n\,\Gamma(5/3)}{2^n\,(2\chi)^{n-1} \Gamma(n+1)\Gamma(5/3-n)} =\frac{\chi}{2}(8+\frac{6x}{\chi})^{2/3}-2\chi\,,
\end{align}
which is compatible with (\ref{Schwarzschildzp}), we find with (\ref{ResumAlpha})
\begin{align}
\zp&\sim\frac{4\chi(1+\mathfrak{c})+(3+\mathfrak{c})x}{\left(8+\frac{6x}{\chi}\right)^{1/3}}+\sum_{m=1}^\infty \frac{x^{m+1} u_m}{(2\chi)^m(m+1)\left(8+\frac{6x}{\chi}\right)^{1/3}}\,,\nonumber\\
&=\frac{4\chi(1+\mathfrak{c})+(3+\mathfrak{c})x}{\left(8+\frac{6x}{\chi}\right)^{1/3}}+\frac{1}{2\chi\left(8+\frac{6x}{\chi}\right)^{1/3}}\sum_{m=1}^\infty \varphi_m\int_0^x dt\, t^m\nonumber
\end{align}
We can therefore formally resum $\zp$, corrected to leading order, in the following manner
\begin{align}
\zp\sim\frac{1}{\left(8+\frac{6x}{\chi}\right)^{1/3}}\left(2(\zh+x)+\frac{1}{2\chi}\int_0^x dt\,\Phi(t)\right)\,.\label{EffectiveProbePosition}
\end{align}
In the classical case (\emph{i.e.} for $\zh=2\chi=\Phi(t)$), this expression reduces to (\ref{Schwarzschildzp}). Similarly, we find with (\ref{ResumEpsilon})
\begin{align}
&\epsilon(x)=-p_{||}(x)\sim\sum_{m=1}^\infty\,\frac{u_m m x^{m-1}}{4\pi(3x+4\chi)(2\chi)^m}=\frac{1}{8\pi \chi (3x+4\chi)}\sum_{m=1}^\infty \varphi_m\,\frac{d}{dx}x^m\,,\nonumber\\
&p_\perp(x)\sim-\sum_{m=1}^\infty\left[\frac{m(m-1) x^{m-2} u_m}{16\pi (2\chi)^m}+\frac{m x^{m-1} u_m}{16 \pi(3x+4\chi)(2\chi)^m}\right]\nonumber\\
&\hspace{1cm}=-\frac{1}{32\pi\chi}\sum_{m=1}^\infty\varphi_m\left(\frac{d^2}{dx^2}+\frac{1}{3x+4\chi}\frac{d}{dx}\right)x^m\,,
\end{align}
which we can formally resum to 
\begin{align}
&\epsilon(x)=-p_{||}(x)\sim\frac{1}{8\pi \chi\, (3x+4\chi)}\,\frac{d}{dx}\Phi(x)\,,\\
&p_\perp(x)\sim-\frac{1}{32\pi\chi}\,\left(\frac{d^2}{dx^2}+\frac{1}{3x+4\chi}\frac{d}{dx}\right)\Phi(x)\,.\label{EffectiveEigenvalues}
\end{align}
Therefore, under the assumptions \emph{(i)} and \emph{(ii)}, we find a closed form expression of the eigenvalues of the energy momentum tensor, corrected to leading order in the deformations: here we have organised corrections in powers of the deformation of the position of the horizon $\mathfrak{c}$ (which we assumed to be small). However, in the same way as discussed in \cite{DelPiano:2024nrl}, the precise details of this small parameter are not important and we could organise the expansions also in other small quantities that characterise the (quantum) deformations of the black hole. In the following we shall understand the symbol $\sim$ to indicate that quantities are correct including linear corrections in such a small parameter.

We also remark that (\ref{EffectiveEigenvalues}) allows to write the fields (\ref{PhysicalSystem}) of the physical system that effectively describes the black hole, in a much more compact form to leading order
{\allowdisplaybreaks
\begin{align}
&p_f\sim -\frac{1}{64\pi\chi}\,\left(\frac{d^2}{dx^2}+\frac{5}{3x+4\chi}\frac{d}{dx}\right)\Phi(x)\,,\nonumber\\
&\rho_f\sim \frac{1}{8\pi \chi\, (3x+4\chi)}\,\frac{d}{dx}\Phi(x)-\frac{1}{64\pi\chi}\left|\left(\frac{d^2}{dx^2}-\frac{3}{3x+4\chi}\frac{d}{dx}\right)\Phi(x)\right|\,,\nonumber\\
&(\vec{\nabla}\phi)^2\sim\text{max}\left\{\frac{1}{32\pi\chi}\,\left(\frac{d^2}{dx^2}-\frac{3}{3x+4\chi}\frac{d}{dx}\right)\Phi(x),0\right\}\,,\nonumber\\
&E^2\sim\text{max}\left\{-\frac{1}{32\pi\chi}\,\left(\frac{d^2}{dx^2}-\frac{3}{3x+4\chi}\frac{d}{dx}\right)\Phi(x),0\right\}\,.\label{MinimalFieldRealisation}
\end{align}}


\subsection{Example: Hayward Black Hole}
In order to illustrate the expansions in Section~\ref{Sect:EnergyMomentum} as well as the iterative solution of the geometry in Section~\ref{Sect:EMTiterativeSol}, we shall briefly discuss the Hayward black hole, whose metric functions are given in (\ref{HorizonHayward}). For this geometry, the expansion coefficients of the components of the energy momentum tensor (\ref{DefEnergyMomentum}) take the form
\begin{align}
&\epsilon_0=-\ppar{0}=\frac{3\gamma(1+\gamma)}{32\pi \chi^2}\,,\hspace{1cm}\epsilon_1=-\ppar{1}=-\frac{9\gamma(1+\gamma)}{32\pi\chi^3}\,,\hspace{1cm}\epsilon_2=-\ppar{2}=\frac{27\gamma(1+\gamma)(5-2\gamma)}{256\pi \chi^4}\,,\nonumber\\
&\epsilon_3=-\ppar{3}=-\frac{27\gamma(1+\gamma)(15-18\gamma+2\gamma^2)}{512\pi \chi^5}\,,\nonumber\\
&\pperp{0}=\frac{3\gamma(2-\gamma)}{32\pi\chi^2}\,,\hspace{2.1cm}\pperp{1}=-\frac{9\gamma(4-5\gamma)}{64\pi \chi^3}\,,\hspace{2.1cm} \pperp{2}=\frac{27\gamma(20-51\gamma+10\gamma^2)}{512\pi\chi^4}\,,\nonumber\\
&\pperp{3}=-\frac{27\gamma(30-132\gamma+76\gamma^2-5\gamma^3)}{512\pi\chi^5}\,.
\end{align}
Notice that all coefficients vanish for $\gamma=0$, in which case the geometry becomes Schwarzschild (and for which the energy momentum tensor vanishes outside of the horizon). Furthermore, as explained in Section~\ref{Sect:EMTiterativeSol}, the coefficients $\{\varphi_n\}_{n\in\mathbb{N}}$ and $\{\psi_n\}_{n\in\mathbb{N}}$ can be expressed in terms of the $\{\epsilon_n,\pperp{n}\}_{n\in\mathbb{N}^*}$: in this fashion we indeed recover the first few coefficients in (\ref{HaywardMetricCoefficients}). Moreover, we also recover the following consistency condition for the position of the horizon
\begin{align}
0=\frac{9\gamma(\zh^2(1+\gamma)^2-4\chi^2)}{128\chi^3\pi(1+\gamma)\zh^2}\,,
\end{align}
which indeed has as solution $\zh=\frac{2\chi}{1+\gamma}$, which is the correct position of the horizon as given in (\ref{HorizonHayward}). Thus, knowledge of the coefficients of the energy momentum tensor outside of the horizon, also fixes the position of the latter.

\begin{figure}[htbp]
\begin{center}
\includegraphics[width=7.5cm]{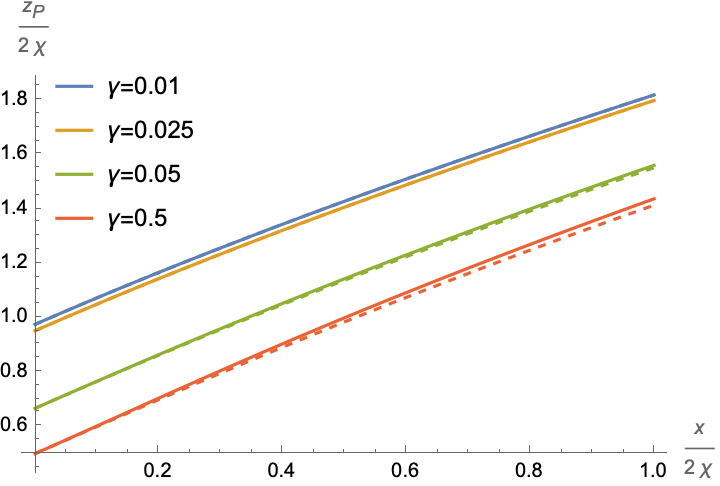}\hspace{1cm}\includegraphics[width=7.5cm]{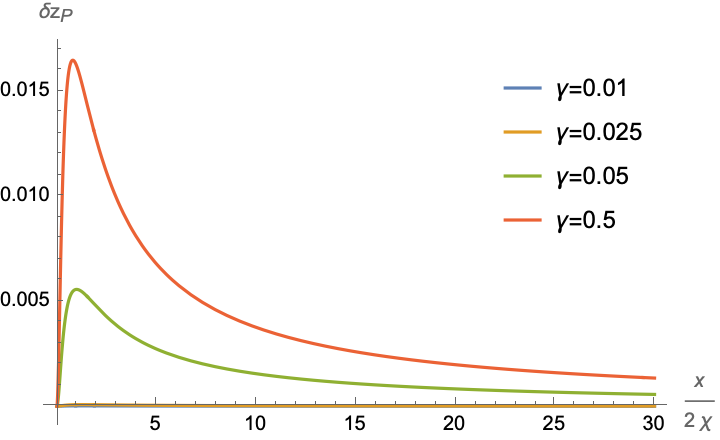}
\end{center}
\caption{\emph{Comparison of $\zp$ in the Hayward space-time (\ref{HorizonHayward}), corrected to leading order (as calculated in (\ref{EffectiveProbePosition})) with a direct numerical computation: the left panel shows $\zp$ for different values of $\gamma$, where the solid lines represent the numerical computations ($\zp^{\text{num}}$) and the dashed lines of matching colour the expression found in (\ref{EffectiveProbePosition}) ($\zp^{\text{eff}}$). For better comparison, the right panel shows the relative error in each case $\delta \zp=\left|\frac{\zp^{\text{num}}-\zp^{\text{eff}}}{\zp^{\text{num}}}\right|$.}}
\label{Fig:HaywardEffectiveZp}
\end{figure}

We can also use the Hayward space-time to test the effective (and partially formal) expressions (\ref{EffectiveProbePosition}) for the position of the probe and (\ref{EffectiveEigenvalues}) for the eigenvalues of the energy momentum tensor. Indeed, with (\ref{HorizonHayward}) we have $\mathfrak{c}=-\frac{\gamma}{1+\gamma}$, which is $\mathcal{O}(\gamma)$, as are the coefficients (\ref{HaywardMetricCoefficients}). Therefore (after rescaling $\{\psi_n\}_{n\in\mathbb{N}}$ by the appropriate factors of $2\chi$ as in (\ref{MassRescaling})), for small enough values of $\gamma$, the Hayward space-time is indeed compatible with (\ref{ScalingApprox}). Furthermore, Figure~\ref{Fig:HaywardEffectiveZp} compares graphically $\zp$ corrected to leading order in (\ref{EffectiveProbePosition}) with the exact (numerical) calculation of $\zp$ as a function of $x$ as defined in (\ref{SeriesExpansionsDeformations}). The plot demonstrates that even for relatively large values of $\gamma$ and at a large region outside of the horizon, the relative error of (\ref{EffectiveProbePosition}) is of the order of less than $1\%$. The plot also indicates that the (relative) error decreases with the distance to the black hole horizon, following the expectation that effects due to the deformation become negligible at large distances to the black hole. 

\begin{figure}[htbp]
\begin{center}
\includegraphics[width=7.5cm]{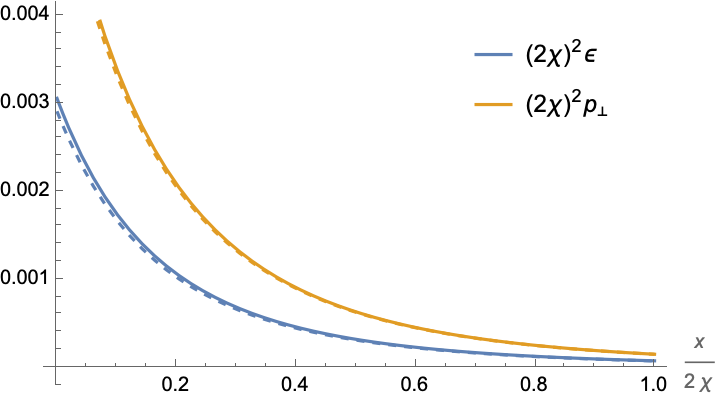}\hspace{1cm}\includegraphics[width=7.5cm]{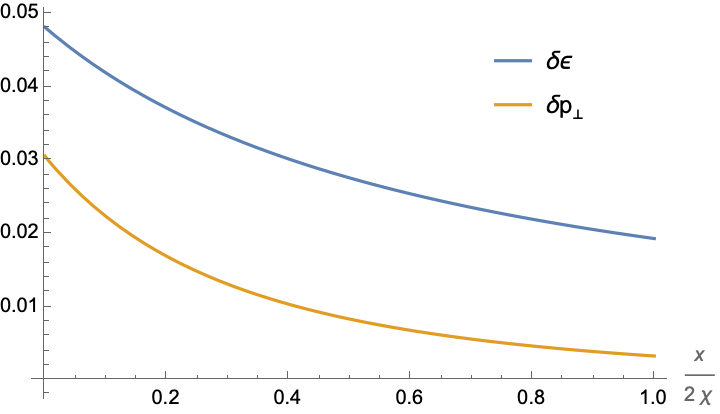}
\end{center}
\caption{\emph{Comparison of the eigenvalues of the energy momentum tensor in the Hayward space-time (\ref{HorizonHayward}), corrected to leading order (as calculated in (\ref{EffectiveEigenvalues})) with a direct numerical computation for $\gamma=0.025$: the left panel shows the eigenvalues $\epsilon(=\!\!-p_{||})$ and $p_\perp$, where the solid lines represent the numerical computations ($\epsilon^{\text{num}}$ and $p_\perp^{\text{num}}$) and the dashed lines of matching colour the expression found in (\ref{EffectiveProbePosition}) ($\epsilon^{\text{eff}}$ and $p_\perp^{\text{eff}}$). The right panel shows the relative error in each case $\delta\epsilon=\left|\frac{\epsilon^{\text{num}}-\epsilon^{\text{eff}}}{\epsilon^{\text{num}}}\right|$ and $\delta p_\perp=\left|\frac{p_\perp^{\text{num}}-p_\perp^{\text{eff}}}{p_\perp^{\text{num}}}\right|$.}}
\label{Fig:HaywardEffectiveEigen}
\end{figure}

\begin{figure}[htbp]
\begin{center}
\includegraphics[width=9.5cm]{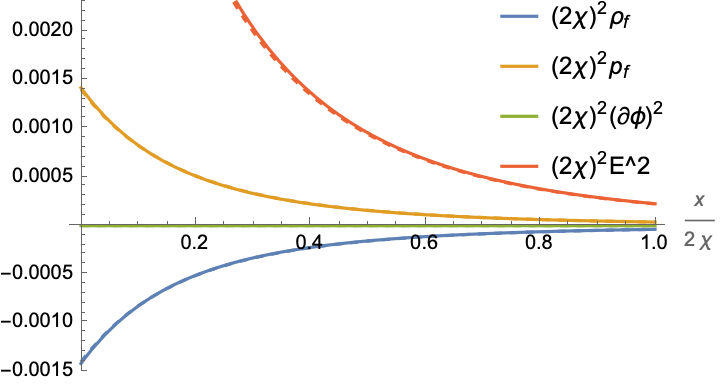}
\end{center}
\caption{\emph{Comparison of the minimal field realisation of the energy momentum tensor in the Hayward space-time (\ref{HorizonHayward}), corrected to leading order (as calculated in (\ref{MinimalFieldRealisation})) with a direct numerical computation for $\gamma=0.025$: the solid lines represent the numerical computations and the dashed lines of matching colour the expression found in (\ref{MinimalFieldRealisation}).}}
\label{Fig:HaywardEffectiveFields}
\end{figure}

Figure~\ref{Fig:HaywardEffectiveEigen} provides a similar comparison between (\ref{EffectiveEigenvalues}) and numerical computations of the eigenvalues $\epsilon(=\!\!-p_{||})$ and $p_\perp$ of the energy momentum tensor as a function of $x$, for a fixed (small) value of $\gamma$. As is evident, also in this case, the agreement is quite good and becomes better further away from the black hole. Finally, Figure~\ref{Fig:HaywardEffectiveFields} shows a comparison between the effective minimal field realisation (\ref{MinimalFieldRealisation}), corrected to first order, and a numerical computation. We note that in the case of the Hayward geometry, the scalar field is absent, while the dispersion relation is $p_f=-\rho_f$ (see also \cite{DAlise:2023hls}).

\section{Conclusions}\label{Sect:Conclusions}
In this work we demonstrate how to unambiguously determine the expansion coefficients of an effective metric description (EMD) of a four-dimensional, spherically symmetric and static black hole,  from observations made from outside of the event horizon. EMDs have been introduced in previous work~\cite{Binetti:2022xdi,DelPiano:2023fiw,DelPiano:2024gvw} as a framework to describe (quantum) deformed black holes in a model-independent way. Indeed, the approach allows to capture universal properties \cite{DAlise:2023hls} of solutions beyond General Relativity (GR) and has been demonstrated to efficiently compute observables in an effective manner~\cite{DelPiano:2024nrl}. At the core of an EMD is a set of expansion coefficients, which encode the metric deformations as functions of a physical quantity, such as a proper time (which we have used in the current work) or proper distances (as used in previous work \cite{Binetti:2022xdi,DelPiano:2023fiw,DelPiano:2024gvw,Li:2023djs,DAlise:2023hls,DelPiano:2024nrl}). Since these coefficients are expected to hold physical significance, they must be accessible by measurements from observers outside of the black hole horizon. Indeed, in this work we have devised a Gedankenexperiment, how to capture these coefficients from the telemetry data sent from probes following a free-falling radial geodesic towards the center of the black hole. Concretely, by measuring a particular redshift factor ($S_1$ in (\ref{Telemetry1})) and a critical emission angle ($S_2$ in (\ref{Telemetry3})) above which photons cannot reach an external observer, we have shown how to iteratively determine all the expansion coefficients of the deformation functions in (\ref{SeriesExpansionsDeformations}). Besides the coefficients $\{\varphi_n,\psi_n\}_{n\in\mathbb{N}}$, this in particular also includes the position of the event horizon $\zh$, which is fixed through the consistency of the telemetry data. Indeed, the equations (\ref{RelAlpha1}), (\ref{HigherAlphaOrders}), (\ref{TelemetryEq1}) and (\ref{TelemetryEq2}) provide iterative (mostly linear) relations that fix all input data for the metric deformations in terms of the coefficients of the telemetry signals. For further use, we have also provided in Section~\ref{Sect:RelProperDistance} relations how to relate our results to an EMD based on the proper distance to the event horizon, which has been used in previous work. Finally, we have showcased and illustrated our computations in the (trivial) case of the Schwarzschild black hole and one of its deformations proposed in the literature~\cite{Hayward:2005gi}.

In a second step, we have established iterative relations between the coefficients of the EMD (based on the proper time of a free-falling observer) and the expansion coefficients of the eigenvalues of the energy momentum tensor. On the one hand, these eigenvalues constitute another type of physical measurements that allow to fully reconstruct the EMD: a probe capable of measuring the energy density $\epsilon$ and pressure components $p_{||}$ and $p_\perp$ (see (\ref{DefEnergyMomentum})) and transmitting them to an observer, would equally allow to fully determine the metric deformations. On the other hand, by mimicking the energy momentum tensor through a (minimal) system of physical fields as in \cite{Boonserm:2015aqa} (\emph{i.e.} a perfect fluid, a massless scalar field and an electric-like field), points towards an effective description of the deformed black hole metric. Indeed, in Section~\ref{Sect:EnergyMomentum}, we have worked out iterative equations that express (the expansion coefficients of) $(\epsilon,p_{||},p_\perp)$ in terms of the coefficients of the EMD. Using the results of \cite{Boonserm:2015aqa}, this provides a direct relation to the physical fields. Furthermore, in the case of a simplified geometry and under the assumption that the deformations are small (in the sense of the scaling relation (\ref{ScalingApprox}), which is similar to the one imposed in \cite{DelPiano:2024nrl}) we provide a closed form expression (\emph{i.e.} not a series expansion) of $(\epsilon,p_{||},p_\perp)$ (see (\ref{EffectiveEigenvalues})) and the physical fields (see (\ref{MinimalFieldRealisation})), corrected to leading order, in terms of the metric functions. Here, following \cite{DelPiano:2024nrl}, we have used the deviation of the horizon position as a parameter of the size of the corrections, however, the result does not depend on the exact details of this definition. We have illustrated these results at the example of the Hayward space-time, for which we find excellent numerical agreement for small enough deformations.

Our results strengthen the effective framework for the description for spherically symmetric and static black holes developed in previous work. They demonstrate that EMDs can indeed \emph{completely} be determined by physical measurements from observers outside of the horizon, thus underlining the physical significance of the effective description. Furthermore, we have established a direct relation between the deformation functions (and their expansion coefficients) that are at the heart of an EMD and a system of physical fields that mimick inhomogeneities entering into effective Einstein equations. While we are not claiming that these equations follow directly from any theory of quantum gravity, they should (at least locally near the horizon) provide equivalent descriptions, provided a geometric description in the form of a metric is still possible. Since we have formulated this equivalent physical system in terms of physical quantities,\footnote{In the current case, we have chosen the proper time of a free-falling observer, which, however, can be changed to other suitable physical quantities if required.} which therefore can be defined unambiguously in any theory of (quantum) gravity, we have demonstrated the universality of the framework. We therefore expect this approach to be useful in the future study of quantum gravity and corrected black hole metrics.

Our discussion in this paper has been limited to general spherically symmetric and static geometries. In the future it will be important to generalise the EMD approach to allow for geometries with non-trivial angular momentum (see \cite{Hohenegger:2024kbg} for such a generalisation in three dimensions). Furthermore, the experiment proposed here is only a Gedankenexperiment, as it requires interaction with the actual black hole in the sense that we have considered a (man-made) probe falling into the black hole. It would be interesting to study, how much of the near-horizon EMD can be recovered by studying (charged) matter falling into the black hole, whose effects could be visible to a distant observer. This would make the discussion directly amenable to astrophysical observations.

\section*{Acknowledgements}
I am very grateful to Mattia Damia Paciarini, Manuel Del Piano, Mikolaj Myszkowski and Francesco Sannino for numerous discussions and exchanges on related topics. I am also thankful to the Quantum Theory Center of the University of Southern Denmark for kind hospitality, while part of this work was being completed.
\appendix
\section{Geodesic Motion}\label{App:GenericGeometry}
In this Appendix we discuss general properties of spherically symmetric and static black hole geometries (in some cases specified to the particular form (\ref{SchwarzschildMetric}) with (\ref{DefDeformation})), which are important for the computations in the main body of this paper: we describe the geodesic motion of test-particles and discuss (the finiteness) of curvature invariants. Many of the concepts reviewed here can be found in textbooks, \emph{e.g.}~\cite{Misner:1973prb,Carroll:2004st}.
\subsection{Geodesic Motion and Proper Time}\label{App:GeodesicProperTime}
Let $x^\mu(\lambda)$ describe the position of a test-particle in 4-dimensional space with the background metric (\ref{SchwarzschildMetric}), which is parametrised by a suitable parameter $\lambda$. The geodesic equations read
\begin{align}
&\frac{du^\mu}{d\lambda}=-{\Gamma^\mu}_{\nu\rho}\,u^\nu\,u^\rho\,,&&\text{with} &&u^\mu=\frac{dx^\mu}{d\lambda}\,,&&\text{and} &&g_{\mu\nu}\,u^\mu\,u^\nu=\zeta\,,\label{GeodesicEquation}
\end{align}
where ${\Gamma^\mu}_{\nu\rho}$ are the connection components of the metric (\ref{SchwarzschildMetric}), $u^\mu$ is the 4-velocity of the test particle and
\begin{align}
\zeta=\left\{\begin{array}{rl}-1 & \text{for a time-like geodesic}\,,\\ 0 & \text{for a light-like geodesic}\,,\\ +1 & \text{for a space-like geodesic}\,.\end{array}\right.
\end{align} 
Furthermore, we assume our coordinate system to be oriented in such a way that the initial conditions are $u^\theta(\lambda=0)=0$, in which case $u^\theta$ remains zero throughout the entire trajectory: the particle remains in a plane of fixed $\theta$ (which we shall choose to be $\theta=\pi/2$). In this case, together with (\ref{SchwarzschildMetric}), the remaining geodesic equations can be reduced to 
\begin{align}
&\frac{du^t}{d\lambda}=-\frac{h'}{h}\,u^t\,u^z\,,&&\frac{du^\varphi}{d\lambda}=-\frac{2}{z}\,u^z\,u^\varphi\,,&&-h\,(u^t)^2+\frac{(u^z)^2}{f}+z^2\,(u^\varphi)^2=\zeta\,,\label{GeodesicEquations}
\end{align}
where $h'(z)=\frac{dh}{dz}(z)$. The first two of these equations can be re-written in the form
\begin{align}
&\frac{d}{d\lambda}\,\log\left(u^t\,h(z)\right)=0\,,&&\text{and} &&\frac{d}{d\lambda}\,\log\left(u^\varphi\,z^2\right)=0\,,
\end{align}
which can be solved by introducing the energy and angular momentum of the particle
\begin{align}
&E=u^t\,h(z)\,,&&\text{and} &&L=z^2\,u^\varphi\,,\label{ConservedQuantities}
\end{align}
which are both conserved along the geodesic motion. 
\subsubsection{Time-like Radial Geodesic}\label{App:TimeLikeGeodesic}
To describe the geodesic of a massive test-particle, which radially falls towards the center of the black hole, we impose
\begin{align}
&\zeta=-1\,,&&\text{and} &&\theta=\frac{\pi}{2}\,,&&\text{and} &&L=0\,,&&\text{and} &&d\varphi=d\theta=0\,.
\end{align}
For further convenience, we shall rescale the time variable $t\to t/E$ in such a way that $u^t\,h(z)=1$ in (\ref{ConservedQuantities}). For the last equation in (\ref{GeodesicEquations}) this implies
\begin{align}
&-h(u^t)^2+\frac{(u^z)^2}{f}=-1\,,&&\text{such that} &&\left(\frac{u^z}{u^t}\right)^2=\frac{f(h(u^t)^2-1)}{(u^t)^2}\,.
\end{align}
Furthermore, we shall choose the parameter $\lambda$ in (\ref{GeodesicEquation}) to be the proper time (measured by this test particle) for falling from (fixed) $z_0$ to $z<z_0$, which is given by
\begin{align}
\tau(z)=-\int_{z_0}^z\sqrt{-ds^2}\,.\label{ProperTime}
\end{align}
Substituting the geodesic equations (\ref{GeodesicEquations}) into the line element (\ref{SchwarzschildMetric}), we obtain
\begin{align}
ds^2=\frac{dz^2}{f}\left(1-f\,h\,\left(\frac{dt}{dz}\right)^2\right)=\frac{dz^2}{f}\left(1-f\,h\,\left(\frac{u^t}{u^z}\right)^2\right)=\frac{dz^2}{f}\left(1-\frac{fh (u^t)^2}{f(h(u^t)^2-1)}\right)\,.
\end{align}
Using furthermore the first equation in (\ref{ConservedQuantities}) we thus obtain
\begin{align}
ds^2=-\frac{h}{f(1-h)}\,dz^2\,,
\end{align}
such that the proper time (\ref{ProperTime}) follows the differential equation
\begin{align}
&\frac{d\tau}{dz}=-\sqrt{\frac{h(z)}{f(z)\,(1-h(z))}}\,,&&\text{with} &&\tau(z_0)=0\,.\label{DiffEqProperTime}
\end{align}
Finally, upon denoting the position of the (external) event horizon by $\zh$, we provide the proper time for the test-particle to fall from $z_0>\zh$ all the way to the horizon
\begin{align}
\tah=-\int_{z_0}^{\zh}\sqrt{\frac{h(z)}{f(z)\,(1-h(z))}}\,dz\,.\label{DiffEqTime}
\end{align}
Notice in particular, if both $f$ and $h$ have simple zeroes at $z=\zh$, we find that the first derivative at the horizon $\frac{d\tau}{dz}\big|_{z=\zh}$ is in general finite.
\subsubsection{Light-like Geodesic}\label{App:LightLikeGeodesic}
To describe the geodesic of a massless test particle, which follows a time-like geodesic, we impose
\begin{align}
&\theta=\frac{\pi}{2}\,,&&L=z^2\,u_\gamma^\varphi\,,&&\zeta=0\,,&&E_\gamma=u^t\,h(z)&&\text{with} &&h(z)\,(u^t)^2=\frac{(u^z)^2}{f(z)}\,,
\end{align}
where $E_\gamma$ and $L$ are independent of the position $z$ of the test particle. Its 4-velocity furthermore satisfies
\begin{align}
&0=-h\,(u^t)^2+\frac{(u^z)^2}{f}+z^2\,(u^\varphi)^2\,,
\end{align}
such that we have in general
\begin{align}
u^\mu=\left(\frac{E_\gamma}{h}\,,E_\gamma\,\sqrt{\frac{f}{h}\,\left(1-\frac{L^2 h}{z^2 E_\gamma^2}\right)}\,,0\,,\frac{L}{z^2}\right)\,.\label{DefFourVelocityPhoton}
\end{align}

\subsection{Finiteness of Curvature Invariants}\label{Sect:CurvatureInvariants}
As discussed in \cite{DelPiano:2023fiw}, an EMD based on the proper distance to the horizon is only consistent under certain conditions on the corresponding expansion coefficients: indeed, since the derivative of $\rho$ (with respect to the radial coordinate $z$) in (\ref{DiffEqDistance}) is singular at the horizon, care has to be taken that curvature invariants such as the Ricci- and Kretschmann scalar remain finite at the horizon. This imposes that the expansions of functions $\Phi$ and $\Psi$ (as in (\ref{SeriesDistance})) only contain even powers. In this appendix we verify if similar conditions also exist for the EMD based on the proper time of a free-falling observer (see also \cite{DamiaPaciarini:2025akr}). Although, under the assumption that both $f$ and $h$ have simple zeroes at $z=\zh$, the first derivative of the proper time at the horizon $\frac{d\tau}{dz}\big|_{z=\zh}$ is in general finite, this is not immediately obvious for the second derivative
\begin{align}
\frac{d^2\tau}{dz^2}\bigg|_{z=\zh}=\frac{1}{2\tah}\lim_{z\to\zh}\left(\frac{h'f-hf'}{f^2(1-h)^2}+\frac{h^2 f'}{f^2(1-h)^2}\right)\,,\label{FormDDer}
\end{align}
If this limit does not exist, the Ricci- and Kretschmann scalar may have singularities for $\tau\to \tah$ (\emph{i.e.} $x\to 0$). However, (\ref{FormDDer}) can be expanded in powers of $x$
\begin{align}
\frac{d^2\tau}{dz^2}=\frac{\left(1+\alpha_1\sqrt{\frac{\alpha_1-\psi_1}{\alpha_1-\varphi_1}}\right)(\psi_1-\varphi_1)}{2(\alpha_1-\psi_1)^2\sqrt{\frac{\alpha_1-\psi_1}{\alpha_1-\varphi_1}} \,x}+\mathcal{O}(x^0)\,,
\end{align}
where we have implicitly assumed $\varphi_1\neq\alpha_1\neq\psi_1$. The singularity is absent for
\begin{align}
&\alpha_1\sqrt{\frac{\alpha_1-\psi_1}{\alpha_1-\varphi_1}}=-1&&\text{or} &&\varphi_1=\psi_1\,.\label{SectStrongerEq}
\end{align}
Notice that (\ref{RelAlpha1}) is a consequence of the first relation, while it implies $\alpha_1=\pm1$ for the second relation. Thus, in general, finiteness of the Ricci- or Kretschmann scalar does not imply conditions on the $\varphi_1$ or $\psi_1$ but rather fixes the coefficient $\alpha_1$ in a way compatible with (\ref{RelAlpha1}).

\section{Photon Captured at Generic Angle}\label{App:FullAngleExpansion}
In Section~\ref{Sect:AngleEmission} we have proposed $S_2$ in (\ref{Telemetry3}) as the second telemetry signal to probe the black hole geometry in the vicinity of the horizon. $S_2$ corresponds to the critical angle, under which photons emitted by the probe can still reach the spacecraft and has the advantage that it does not require to know the photon trajectory $\zgam$ at intermediate radial positions $\zp<\zgam<\zs$. In this Appendix we briefly argue that there are in fact also other possible ways to extract suitable telemetry data. Indeed, while the angle (\ref{FormAngleSpacecraft}) under which the photon emitted in Section~\ref{Sect:AngleEmission} can be measured by the spacecraft, the integral presentation notably requires to know the metric functions $f$ and $h$ along the entire path of the integral. In order to still make use of this measurement, we therefore prefer an expression that is only sensitive to information at $\zp$, \emph{i.e.} the location where the photon was originally emitted. To this end, we can for example compute the derivative of $\varphi$ with respect to $x$, \emph{i.e.} the (proper time measured by the probe) until the probe crosses the horizon
\begin{align}
&\frac{d\varphi}{dx}=-\frac{d\zp}{dx}\,\frac{\tan\ang}{\zp\,\sqrt{f(\zp)}}+\frac{d\imp}{dx}\,\int_{\zp}^{\zs}\,dz\,G\,,&&\text{with}&&G(z,\imp)=\frac{\partial}{\partial\imp}\,\left(\frac{\imp}{z^2\sqrt{\frac{f}{h}\,\left(1-\frac{\imp^2 h}{z^2}\right)}}\right)\,.\label{FunctionalFormAngle}
\end{align}
The second term in this expression is still an integral. However, we can find further information about this term by considering the derivative of $\varphi$ in (\ref{FormAngleSpacecraft}) with respect to $\ang$.\footnote{This function can be obtained by studying the emission of a second photon from the probe at a slightly modified angle $\ang'=\ang+\delta\ang$, with $\delta\ang\ll 1$.} In this case, the change of the angle $\varphi$ in (\ref{FormAngleSpacecraft}) takes the form
\begin{align}
\frac{\partial\varphi}{\partial\ang}=\frac{\zp\,\cos\ang}{\sqrt{h(\zp)}}\,\int_{\zp}^{\zs}\,dz\,G\,,
\end{align}
such that (\ref{FunctionalFormAngle}) can be expressed in the form
\begin{align}
\frac{d\varphi}{dx}=-\frac{d\zp}{dx}\,\frac{\tan\ang}{\zp\,\sqrt{f(\zp)}}+\left[\tan\ang\left(\frac{1}{\zp}\,\frac{d\zp}{dx}-\frac{1}{2h(\zp)}\frac{dh(\zp)}{dx}\right)+\frac{d\ang}{dx}\right]\,\frac{\partial\varphi}{\partial\ang}\,.\label{TelemetryEq4}
\end{align}
which is equivalent to 
\begin{align}
\frac{1}{\tan\ang}\left(\frac{d\varphi}{dx}-\frac{\partial\varphi}{\partial\ang}\,\frac{d\ang}{dx}\right)=\frac{1}{\zp}\,\frac{d\zp}{dx}\left(\frac{\partial\varphi}{\partial\ang}-\frac{1}{\sqrt{f}}\right)-\frac{1}{2h(\zp)}\,\frac{d h(\zp)}{dx}\,\frac{\partial\varphi}{\partial\ang}\,.\label{DetAngleExpand}
\end{align}
The quantities in this expression are either measured by the spacecraft ($\frac{d\varphi}{dx}$, $\frac{\partial\varphi}{\partial\ang}$), transmitted by the probe ($\frac{d\ang}{dx}$) or contain unknown coefficients of the metric deformations we wish to determine ($\zp$, $\sqrt{f}$). Thus, upon expanding (\ref{DetAngleExpand}) in powers of $x$, allows to establish relations between the coefficients $\{\alpha_n\}$ and $\{\varphi_n\}$ in terms of known coefficients. 
\section{Series Expansions}\label{App:PowerInversion}
In this Appendix we work out series expansions of a number of quantities, which are useful in the main body of the paper, namely inverse powers of the probe position $\frac{1}{\zp^N(x)}$ and the derivative $\frac{d\tau}{dz}$ of the proper time of the the probe as functions of $x$ (defined in (\ref{SeriesExpansionsDeformations})).
\subsection{Expansion of $\frac{1}{\zp^N(x)}$}
In this appendix, starting from an expansion of $z$ as a function of $x$ (see the first equation in (\ref{SeriesExpansionsDeformations})), we determine the series expansion
\begin{align}
Q(x)=\frac{1}{\zp^N(x)}=\frac{1}{\zh^N}+\sum_{n=1}^\infty \mathfrak{q}^{(N)}_n\,x^n\,,\label{DerivativeIdentity}
\end{align}
for $N\in\mathbb{N}$ an integer and  $\mathfrak{q}^{(N)}_n\in\mathbb{R}$. To this end, we consider
\begin{align}
\zp(x)\,Q'(x)=-N Q(x)\,\frac{d\zp}{dx}\,,
\end{align}
which leads to the relation 
\begin{align}
\left(\zh+\sum_{n=1}^\infty \alpha_n x^n\right)\left(\sum_{m=1}^\infty m\, \mathfrak{q}^{(N)}_m\,x^{m-1}\right)=-N \left(\frac{1}{\zh^N}+\sum_{n=1}^\infty \mathfrak{q}^{(N)}_n\,x^n\right)\left(\sum_{m=1}^\infty m \,\alpha_m x^{m-1}\right)\,.
\end{align}
Combining the summations on the left and right hand side, we obtain
\begin{align}
&0=\left(\zh\,\mathfrak{q}^{(N)}_1+\frac{N\alpha_1}{\zh^N}\right)\nonumber\\
&+\sum_{p=1}^\infty x^p \bigg[(p+1)\,\left(\zh \mathfrak{q}^{(N)}_{p+1}+\frac{N\alpha_{p+1}}{\zh^N}\right)+\sum_{n=1}^p(p-n+1)(\alpha_n\, \mathfrak{q}^{(N)}_{p-n+1}+N\alpha_{p-n+1}\mathfrak{q}^{(N)}_n)\bigg]\,.
\end{align}
We can thus read off the coefficients 
\begin{align}
&\mathfrak{q}^{(N)}_1=-\frac{N\alpha_1}{\zh^{N+1}}\,,&&\text{and} &&\mathfrak{q}^{(N)}_{p}=-\frac{1}{p\zh}\left[\frac{Np\alpha_{p}}{\zh^N}+\sum_{n=1}^{p-1}(p-n)(\alpha_n\, \mathfrak{q}^{(N)}_{p-n}+N\alpha_{p-n}\mathfrak{q}^{(N)}_n)\right]\,,\hspace{0.2cm}\forall p>1\,,\label{ExpansionCoPowers}
\end{align}
where the last relation allows to iteratively calculate all $\mathfrak{q}^{(N)}_p$.

\subsection{Expansion of $\tau'(\zp)$}
From the expansion of $\zp$ as a function of $x$ in (\ref{SeriesExpansionsDeformations}), we can also determine the expansion of $\tau'=\frac{d\tau}{d\zp}$. Indeed, we write the latter as
\begin{align}
\frac{d\tau}{d\zp}=\sum_{n=0}^\infty\mathfrak{p}_n\,x^n\,,
\end{align}
for $x=\tah-\tau$. To determine the coefficients $\mathfrak{p}_n$, we impose
\begin{align}
-1=\left(\sum_{n=0}^\infty\mathfrak{p}_nx^n\right)\left(\sum_{m=1}^\infty m\alpha_m x^{m-1}\right)=\left(\mathfrak{p}_0+\sum_{n=1}^\infty\mathfrak{p}_nx^n\right)\left(\alpha_1+\sum_{m=1}^\infty (m+1)\alpha_{m+1} x^{m}\right)\,.
\end{align}
Developing the product of the two summations, we find 
\begin{align}
-1=\mathfrak{p}_0\,\alpha_1+\sum_{p=1}^\infty x^p \left(\alpha_1\mathfrak{p}_p+(p+1)\,\mathfrak{p}_0\,\alpha_{p+1}+\sum_{n=1}^{p-1}(p-n+1)\,\mathfrak{p}_n\,\alpha_{p-n+1}\right)\,.
\end{align}
We can thus iteratively read off the coefficients 
\begin{align}
&\mathfrak{p}_0=-\frac{1}{\alpha_1}\,,&&\text{and} &&
\mathfrak{p}_p=-\frac{(p+1)}{\alpha_1}\,\mathfrak{p}_0\,\alpha_{p+1}-\frac{1}{\alpha_1}\,\sum_{n=1}^{p-1}(p-n+1)\,\mathfrak{p}_n\,\alpha_{p-n+1}\hspace{0.5cm}\forall p>1\,.\label{ExpansionCoDer}
\end{align}




\printbibliography

\end{document}